\pdfoutput=1
\RequirePackage{ifpdf}
\ifpdf 
\documentclass[pdftex]{sigma}
\else
\documentclass{sigma}
\fi

\newcommand{\bfx}{{\mathbf x}}
\newcommand{\bfxp}{{{\mathbf x}^\prime}}
\newcommand{\wbfx}{\widehat{\mathbf x}}
\newcommand{\wbfxp}{{\widehat{\mathbf x}^\prime}}
\newcommand{\N}{{\mathbb N}}

\newcommand{\Q}{{\mathbb Q}}
\newcommand{\R}{{\mathbb R}}
\newcommand{\Si}{{\mathbb S}}

\newcommand{\li}{{\mathfrak l}}

\newcommand{\Z}{{\mathbb Z}}
\newcommand{\C}{{\mathbb C}}

\newcommand{\hii}{{\mathfrak h}}

\newcommand{\mcg}{{\mathcal G}}

\numberwithin{equation}{section}

\begin{document}

\allowdisplaybreaks

\renewcommand{\PaperNumber}{042}

\FirstPageHeading

\ShortArticleName{Fourier, Gegenbauer and Jacobi Expansions}

\ArticleName{Fourier, Gegenbauer and Jacobi Expansions\\ for a Power-Law Fundamental Solution\\ of the Polyharmonic Equation\\ and Polyspherical Addition Theorems}

\Author{Howard S.~COHL}

\AuthorNameForHeading{H.S.~Cohl}

\Address{Applied and Computational Mathematics Division,
National Institute of Standards\\ and Technology,
Gaithersburg, MD, 20899-8910, USA}
\Email{\href{mailto:howard.cohl@nist.gov}{howard.cohl@nist.gov}}
\URLaddress{\url{http://hcohl.sdf.org}}

\ArticleDates{Received November 29, 2012, in f\/inal form May 28, 2013; Published online June 05, 2013}

\Abstract{We develop complex Jacobi, Gegenbauer
and Chebyshev polynomial expansions for the kernels associated with
power-law fundamental solutions of the polyharmonic equation
on $d$-dimensional Euclidean space.
From these series representations we derive Fourier expansions in
certain rotationally-invariant coordinate
systems and Gegenbauer polynomial expansions in Vilenkin's polyspherical
coordinates.  We compare both of these
expansions to generate addition theorems
for the azimuthal Fourier coef\/f\/icients.}


\Classification{35A08; 31B30; 31C12; 33C05; 42A16}

\section{Introduction}
\label{Introduction}

We have developed a technique for constructing addition theorems for the
azimuthal Fourier coef\/f\/icients of fundamental solutions for linear
homogeneous partial
dif\/ferential equations on $d$-dimensional isotropic Riemannian manifolds.
For a fundamental solution, we construct azi\-muthal Fourier expansions
and compare them with eigenfunction expansions in rotationally-invariant coordinate systems.
The construction of eigenfunction expansions for fundamental solutions
in separable coordinate systems is in general non-trivial.

In connection with the Laplace operator, we have already constructed some
of these addition theorems on $\R^3$ \cite{Cohlerratum12,CRTB,CTRS}, where
we have treated the spherical, cylindrical, oblate spheroidal, prolate spheroidal,
parabolic, bispherical and toroidal coordinate systems.
One may construct addition theorems in this manner in any rotationally-invariant
coordinate system which yields solutions through separation of variables for the
Laplace equation.  In a similar setting, addition theorems may be generated for
other inhomogeneous linear partial dif\/ferential equations, such as for the Helmholtz,
wave and heat equations in arbitrary dimensions.  Furthermore an extension of this
concept is possible when working with linear partial dif\/ferential operators on
Riemannian manifolds, such as for the
Laplace--Beltrami operator (see 
\cite[\S~5.1]{CohlKalII}).
Once a Fourier expansion
for a fundamental solution is obtained for a partial dif\/ferential operator on
a Riemannian mani\-fold, one must construct eigenfunction expansions
for a fundamental solution corresponding to that operator and identify
those nested multi-summation and multi-integration eigenfunction
expansions which correspond to the Fourier coef\/f\/icients for that operator.

In this paper, we apply this technique to generate addition theorems
from eigenfunction expansions for a fundamental solution of the polyharmonic
operator on $d$-dimensional Euclidean space $\R^d$ in
Vilenkin--Kuznetsov--Smorodinski{\u\i} polyspherical
coordinate systems
\cite[\S~9.5]{Vilen}, \cite{VilKuzSmor},
\cite[\S~10.5]{VilenkinKlimyk2} (hereafter Vilenkin).
We have computed azimuthal Fourier expansions for a fundamental solution
of this operator, as well as the corresponding eigenfunction
expansions in polyspherical coordinates.  In each case, the comparison
of these two expansions yields new addition theorems for the azimuthal
Fourier coef\/f\/icients.

The main results of this paper are connected with closed-form
expressions of hypergeometric
orthogonal polynomial expansions for a power-law fundamental solution
of the polyharmonic equation on $d$-dimensional Euclidean space.  These
expansions are the fundamental building blocks for algorithms to compute
the solution of inhomogeneous linear polyharmonic boundary value problems
through convolution with the source distribution.
These problems are ubiquitous in physics and engineering and include
elasticity, electrostatics, magnetostatics, quantum direct and exchange Coulomb
interactions, Newtonian gravity, and potential f\/luid and heat f\/low, just to
name a few.  The expansions presented in this paper are crucial for obtaining analytic
polyharmonic solutions and for numerical algorithms where expansions are needed
for so-called ``fast algorithms''.  Applications of generalized Hopf coordinates,
one of the polyspherical coordinate systems we study in this paper include
particle physics~\cite{LinYang},
quantum f\/ield theory~\cite{Izquierdoetal} and cosmology~\cite{Lakeetal}.
Furthermore, expansions in hyperspherical coordinate systems have
many applications including general atomic
multibody theory (see~\cite{FanoRau,Lin} and references therein).

From a global analytic partial dif\/ferential equation perspective,
the most important results of this paper
are contained in Corollaries~\ref{MYCOROLLARY} and~\ref{MYOTHERCOROLLARY}.  These formulae represent multipole and
azimuthal Fourier decompositions for arbitrary powers of the Euclidean distance
between two points. They can be used to analytically and
numerically solve, in a rapidly convergent fashion, the inhomogeneous linear
polyharmonic equation for isolated non-axisymmetric source distributions.
Rapid convergence of the Fourier expansions is provided by the fact that
for each azimuthal mode there corresponds an inf\/inite number of meridional
modes, which are all summed over in our expansions.
Furthermore, if pure trigonometric azimuthal dependence exists for a particular
source distribution in the inhomogeneous partial dif\/ferential equation,
then Corollary~\ref{MYOTHERCOROLLARY} provides a solution in a f\/inite number of terms.
In the case of an axisymmetric source distribution, the inhomogeneous polyharmonic
solution is obtained from a single $m=0$ term in the expansion.

Corollary~\ref{MYCOROLLARY} is powerful in that it provides a mechanism for
determining the multipole moments associated with unrestricted powers of the
distance between two points. This is in contrast with the Gegenbauer generating
function (which is generalized by Theorem~\ref{geneneralizationofgeneratingfuncitonforgegenbauerpolyJAC}
and Corollary~\ref{geneneralizationofgeneratingfuncitonforgegenbauerpoly})
which provides with the addition theorem for hyperspherical harmonics~(\ref{addthmhypsph})
a multipole expansion for this kernel only for powers of the distance
given
by $\nu=2-d$, for $d=3,4,5,\ldots$.

From a special function theoretic perspective, the new results presented
in this paper are Theorem~\ref{geneneralizationofgeneratingfuncitonforgegenbauerpolyJAC} and Corollary~\ref{geneneralizationofgeneratingfuncitonforgegenbauerpoly}. These series
expansions represent fundamental generalizations of Heine's formula
\cite[(14.28.2)]{NIST}, Gegenbauer's generating function
\cite[(18.12.4)]{NIST}, and Heine's reciprocal square root identity
\cite[(3.11)]{CohlDominici}.  Formula~(\ref{QJACQLEG}) which provides a connection
between the symmetric Jacobi function of the second kind and the associated Legendre
function of the second kind, is also interesting. As far as the author
is aware, this has not previously appeared in the literature.  The addition
theorems for associated Legendre functions
given by Theorems~\ref{additiontheoremrddgt3standard} and~\ref{additiontheoremr2q},
and their
Corollaries~\ref{additiontheoremba},
\ref{additiontheoremb2a},
\ref{additiontheoremca2}, also appear to be new.
This paper only uses eigenfunction expansions for a power-law
fundamental solution of the polyharmonic equation in Vilenkin's polyspherical
coordinates to obtain new addition theorems for the azimuthal Fourier coef\/f\/icients.
We have only treated two dif\/ferent types of Vilenkin's polyspherical
coordinates.  In higher dimensions, many more types may be considered.
The azimuthal Fourier expansion presented in this paper, Corollary~\ref{MYOTHERCOROLLARY}, can be used to provide
new addition theorems for the azimuthal Fourier coef\/f\/icients
in every rotationally-invariant
coordinate system which separates the polyharmonic equation
on~$\R^d$.  These rotationally-invariant coordinate systems
include those of cylindrical, parabolic, and cyclidic type.

In this paper, we take advantage of previously derived closed-form expressions for
the separated eigenfunctions in Vilenkin's polyspherical coordinates
(found in \cite{IPSWb,VilenkinKlimyk2} and elsewhere) to derive addition
theorems from a power-law fundamental solution
of the polyharmonic operator in Euclidean space~$\R^d$.
These addition theorems separate the complicated
geometrically-relevant quantity (the azimuthal Fourier coef\/f\/icients)
into functions of the individual variables in the problem.
We study all dimensions $d\ge 3$ with an emphasis on the
simplicity/explicitness of the low-dimensional examples.

This paper is organized as follows.
In Section~\ref{Prelude}, we describe the kernels associated with a~fundamental
solution of the polyharmonic equation on Euclidean space~$\R^d$
for $d\ge 2$ and introduce rotationally-invariant coordinate systems.
In Section~\ref{Jacobi}, we prove several new theorems associated
with Jacobi, Gegenbauer, and Chebyshev polynomial expansions for the
kernels associated with power-law fundamental solutions
of the polyharmonic equation on~$\R^d$.
In Section~\ref{Multisummationadditiontheorems}, we derive
and discuss new addition theorems in Vilenkin's
polyspherical coordinates for the azimuthal Fourier coef\/f\/icients
of a fundamental solution for the polyharmonic equation
on~$\R^d$.
In Appendix~\ref{Specialfunctionsandorthogonalpolynomials}, we summarize
the def\/initions and properties of the special functions and ortho\-go\-nal
polynomials that we use.
In Appendix~\ref{MethodofTrees}, we review Vilenkin's
polyspherical coordinates and the corresponding normalized
hyperspherical harmonics.

Throughout this paper we rely on the following def\/initions.
Let $a_1,a_2,a_3,\ldots\in\C$, with $\C$ being the set of
complex numbers.  If $i,j\in\Z$ and $j<i$, then
$\sum\limits_{n=i}^{j}a_n=0$ and $\prod\limits_{n=i}^ja_n=1$.
The set of natural numbers is given by $\N:=\{1,2,3,\ldots\}$, the set
$\N_0:=\{0,1,2,\ldots\}=\N\cup\{0\}$, and $\Z:=\{0,\pm 1,\pm 2,\ldots\}$.
The sets $\Q$ and $\R$ represents the rational and real numbers respectively.
For $d\in\N$, we denote by $\R^d$, the f\/inite-dimensional vector space,
$d$-dimensional Euclidean space.
Furthermore, if $\bfx,\bfxp\in\R^d$ then
the Euclidean inner product $(\cdot,\cdot):\R^d\times\R^d\to\R$
def\/ined by
\begin{gather}
(\bfx,\bfxp):=x_1x_1^\prime+\cdots+x_dx_d^\prime,
\label{eucinnerprod}
\end{gather}
induces a norm (the Euclidean norm) $\|\cdot\|:\R^d\to[0,\infty)$,
on $\R^d$, given by $\|\bfx\|:=\sqrt{(\bfx,\bfx)}$.

  Please see Appendix~\ref{Specialfunctionsandorthogonalpolynomials} for all
notations used in this paper for special functions and orthogonal polynomials.

\section{Fundamental solution of the polyharmonic equation\\ in
rotationally-invariant and polyspherical coordinate systems}
\label{Prelude}

In Euclidean space $\R^d$, let the Laplacian
operator $\Delta:C^p(\R^d)\to C^{p-2}(\R^d)$ for $p\ge 2$
be def\/ined by
$\Delta:=
\frac{\partial^2}{\partial x_1^2}+\dots+\frac{\partial^2}{\partial x_d^2}$.
If $\Phi:\R^d\to\R$ satisf\/ies the polyharmonic equation given by
\begin{gather}
(-\Delta)^k\Phi(\bfx)=0,
\label{polyharmoniceq}
\end{gather}
where $\bfx\in\R^d$, $k\in\N$ and $\Phi\in C^{2k}(\R^d)$,
then $\Phi$ is called polyharmonic.
We use the nonnegative Laplacian $-\Delta\ge 0$.
The inhomogeneous polyharmonic equation is given by
\begin{gather}
(-\Delta)^k\Phi(\bfx)=\rho(\bfx),
\label{polyh}
\end{gather}
where we take $\rho$ to be an integrable function so that
a solution to (\ref{polyh}) exists.  A fundamental solution for the
polyharmonic equation on $\R^d$ is a function
${\mcg}_k^d:(\R^d\times\R^d)\setminus\{(\bfx,\bfx):\bfx\in\R^d\}\to\R$
which satisf\/ies the equation
\begin{gather*}
(-\Delta)^k{\mcg}_k^d({\bfx},{\bfx}^\prime)=\delta({\bfx}-{\bfx}^\prime),
\end{gather*}
where $\delta$ is the Dirac delta function (generalized function/distribution)
and $\bfxp\in\R^d$.
Note that this equation is satisf\/ied in the sense of distributions.

A fundamental solution of the polyharmonic equation
is given as follows
(see for instance
\cite{Boyl}, \cite[p.~202]{GelfandShilov}, \cite[p.~45]{Schw}).
Let $d,k\in\N$.  Def\/ine
$\mcg_k^d : (\R^d\times\R^d)\setminus\{(\bfx,\bfx) : \bfx\in\R^d\}\to\R$
by
\begin{gather}
\mcg_k^d({\bfx},{\bfx}^\prime):=
 \begin{cases}
{\displaystyle \frac{(-1)^{k+d/2+1}\ \|{\bfx}-{\bfx}^\prime\|^{2k-d}}
{(k-1)! \left(k-d/2\right)! 2^{2k-1}\pi^{d/2}}
\left(\log\|{\bfx}-{\bfx}^\prime\|-\beta_{k-d/2,d}\right)}\\
\hspace{7.65cm} \mathrm{if}\  d\ \mathrm{even},\ k\ge d/2,\\
{\displaystyle \frac{\Gamma(d/2-k)\|{\bfx}-{\bfx}^\prime\|^{2k-d}}
{(k-1)! 2^{2k}\pi^{d/2}}} \hspace{3.62cm} \mathrm{otherwise},
\end{cases}
\label{greenpoly}
\end{gather}
where $\beta_{p,d}\in\Q$ is def\/ined as
$\beta_{p,d}:=\frac12\left[H_p+H_{d/2+p-1}-H_{d/2-1} \right]$,
with $H_j\in\Q$ being the $j$th harmonic number
$H_j:=\sum\limits_{i=1}^j\frac1i$.
The gamma function
$\Gamma:\C\setminus{-}\N_0\to\C$, is a natural generalization
of the factorial function.
Concerning the logarithmic contribution for $d$ even, $k\ge d/2$,
the polynomial $\|\bfx-\bfxp\|^{2k-d}$ is polyharmonic, so any choice for
the constant $\beta_{p,d}$ is valid.
Our choice for this constant is given such that
$-\Delta \mcg_k^d=\mcg_{k-1}^d$ is satisf\/ied for all $k> d/2$,
and that for $k=d/2$, the constant vanishes.
Note that a solution of the inhomogeneous polyharmonic equation~(\ref{polyh}) is obtained from $\mcg_k^d$ via a convolution.

\subsection{Fundamental solution decompositions \\ in rotationally-invariant coordinate systems}

In this paper we restrict our
attention to separable rotationally-invariant coordinate systems for the polyharmonic
equation on $\R^d$ which are given by
\begin{gather}
\begin{split}
&x_1= R(\xi_1,\ldots,\xi_{d-1})\cos\phi,\\
&x_2 = R(\xi_1,\ldots,\xi_{d-1})\sin\phi,\\
&x_3 = x_3(\xi_1,\ldots,\xi_{d-1}),\\
& \cdots\cdots\cdots\cdots\cdots\cdots\cdots\cdots \\
&x_d = x_d(\xi_1,\ldots,\xi_{d-1}).
\end{split}
\label{rotatioanallyinvariant}
\end{gather}
These coordinate systems are described by $d$-coordinates: an angle
$\phi\in[0,2\pi)$ plus $(d-1)$-curvilinear coordinates $(\xi_1,\ldots,\xi_{d-1})$.
Rotationally-invariant coordinate systems parametrize points on the $(d-1)$-dimensional
half-hyperplane given by $\phi=$ const and $R\ge 0$ using the curvilinear coordinates
$(\xi_1,\ldots,\xi_{d-1})$.
A separable rotationally-invariant coordinate system transforms the polyharmonic equation
into a set of $d$-uncoupled ordinary dif\/ferential equations with separation constants
$m\in\Z$ and $k_j\in\R$ for $1\le j\le d-2$.
For a separable rotationally-invariant
coordinate system, this uncoupling
is accomplished, in general, by assuming a product solution
to (\ref{polyharmoniceq}) of the form
$\Phi(\bfx)=e^{im\phi} {\mathcal R}(\xi_1,\ldots,\xi_{d-1})\prod\limits_{i=1}^{d-1}
A_i(\xi_i,m,k_1,\ldots,k_{d-2})$,
where the properties of the functions
${\mathcal R}$ and $A_i$, for $1\le i\le d-1$, and the constants $k_j$ for
$1\le j\le d-2$, depend on the specif\/ic separable rotationally-invariant
coordinate system in question.
Separable coordinate systems are divided into two dif\/ferent classes,
those which are simply separable (${\mathcal R}= \operatorname{const}$), and those which are ${\mathcal R}$-separable (see~\cite{Miller}).

The Euclidean distance between two points $\bfx,\bfxp\in\R^d$ expressed in the
rotationally-invariant coordinate system described
in~(\ref{rotatioanallyinvariant}) is
\begin{gather}
  \|\bfx-\bfxp\|=\sqrt{2RR^\prime}
\left[\chi-\cos(\phi-\phi^\prime)\right]^{1/2},
\label{algebraicdist}
\end{gather}
where the toroidal parameter $\chi$ is
\begin{gather}
\chi:=\frac{R^{2}+{R^\prime}^2
+{\sum\limits_{i=3}^d(x_i-x_i^\prime)^2}
}
{2RR^\prime}.
\label{toroidalparameter}
\end{gather}
The hypersurfaces $\chi=$ const are independent of
coordinate system and represent hypertori of revolution.

{\sloppy We now rewrite (\ref{greenpoly}) in terms of
the rotationally-invariant coordinate system (\ref{rotatioanallyinvariant}).
From~(\ref{greenpoly}) we see that, apart from multiplicative constants,
the expression
$\li_k^d : (\R^d\times\R^d)\setminus \{(\bfx,\bfx) : \bfx \in \R^d\} \to \R$
of a fundamental solution for the polyharmonic equation
on $\R^d$ for $d$ even, $k\ge d/2$, is given by
\begin{gather*}
\li_k^d(\bfx,\bfxp):=\|\bfx-\bfxp\|^{2k-d}\left(\log\|\bfx-\bfxp\|-\beta_{k-d/2,d}\right).
\end{gather*}
By expressing $\li_k^d$ in a rotationally-invariant coordinate system
(\ref{rotatioanallyinvariant}) we obtain
\begin{gather}
 \li_k^d(\bfx,\bfxp)=\left(2RR^\prime\right)^p\left[\frac12\log
\left(2RR^\prime\right)-\beta_{p,d}\right]
\left[\chi-\cos(\phi-\phi^\prime) \right]^p\nonumber\\
\hphantom{\li_k^d(\bfx,\bfxp)=}{}
+\frac12\left(2RR^\prime\right)^p
\left[\chi-\cos(\phi-\phi^\prime) \right]^p
\log\left[\chi-\cos(\phi-\phi^\prime) \right],
\label{logfourseries}
\end{gather}
where $p=k-d/2\in\N_0$.
Similarly, when working on an even-dimensional Euclidean space
$\R^d$ with $1\le k \le d/2-1$, a fundamental solution of the polyharmonic
equation $\hii_k^d:(\R^d\times\R^d)\setminus\{(\bfx,\bfx):\bfx\in\R^d\}
\to(0,\infty)$
is
\[
\hii_k^d(\bfx,\bfxp):=\|\bfx-\bfxp\|^{2k-d}.
\]
By expressing $\hii_k^d$ in a rotationally-invariant coordinate system we obtain
\begin{gather}
\hii_k^d(\bfx,\bfxp)=\left(2RR^\prime\right)^{-q}
\left[\chi-\cos(\phi-\phi^\prime) \right]^{-q},
\label{unlogfourseries}
\end{gather}
where $q=2k-d\in\N$.

}

Examining (\ref{logfourseries}) and (\ref{unlogfourseries}), we see that for
computation of Fourier expansions
about the azimuthal separation
angle $(\phi-\phi^\prime)$ of $\li_k^d$ and $\hii_k^d$, all that is required is to compute
the Fourier cosine series for the following three functions
$f_\chi,h_\chi:\R\to(0,\infty)$ and $g_\chi:\R\to\R$ def\/ined as
\begin{gather*}
 f_\chi(\psi):=\left(\chi-\cos\psi\right)^p,\qquad
 g_\chi(\psi):=\left(\chi-\cos\psi\right)^p\log\left(\chi-\cos\psi\right),\\
 h_\chi(\psi):=\left(\chi-\cos\psi\right)^{-q},
\end{gather*}
where $p\in\N_0$, $q\in\N$ and $\chi>1$ is a f\/ixed parameter.

The Fourier series of $f_\chi$
is given in~\cite{CohlDominici}
(cf.~(4.4) therein)\footnote{We have used
Whipple's formula (\ref{whipple}) in (\ref{integer})
and (\ref{azimuthalfourierseriesofoneoverq})
to convert the associated Legendre functions of the second
kind $Q_\nu^\mu$
appearing in~\cite{CohlDominici} to
associated Legendre functions of the f\/irst
kind $P_\nu^\mu$.}, namely for $p\in\N_0$,
\begin{gather}
(z-x)^p=(z^2-1)^{p/2}\sum_{n=0}^p
\frac{\epsilon_n (-p)_n(p-n)!}{(p+n)!}
P_p^n\left(\frac{z}{\sqrt{z^2-1}}\right)
T_n(x),
\label{integer}
\end{gather}
where $\epsilon_n\in\{1,2\}$ is the Neumann factor
def\/ined by $\epsilon_n:=2-\delta_{n,0}$,
$\delta_{n,0}\in\{0,1\}$ is the Kronecker delta.
The Fourier series of $h_\chi$ is given in
\cite[(4.5)]{CohlDominici},
namely for $p\in\N$,
\begin{gather}
\frac{1}{(z-x)^p}=
\frac{(z^2-1)^{-p/2}}{(p-1)!}
\sum_{n=0}^\infty\epsilon_n (n+p-1)! P_{p-1}^{-n}
\left(\frac{z}{\sqrt{z^2-1}}\right)T_n(x).
\label{azimuthalfourierseriesofoneoverq}
\end{gather}
In order to compute Fourier expansion of $\li_k^d$ (\ref{logfourseries})
in separable rotationally-invariant coordinate systems, all that remains is to determine
the Fourier series of~$g_\chi$ (see~\cite{CohlthesisII}).
A discussion of Fourier cosine expansions for a logarithmic fundamental solution
of the polyharmonic equation on~$\R^d$ (from~$g_\chi$) can be found in~\cite{Cohl12log}.  The corresponding Gegenbauer polynomial expansions
for a logarithmic
fundamental solution of the polyharmonic equation on $\R^d$ can be found in~\cite{CohlthesisII}.

\section{Jacobi polynomial and limiting expansions\\ for the Euler kernel}
\label{Jacobi}

In this section we derive Jacobi, Gegenbauer and Chebyshev polynomial
of the f\/irst kind series expansions of the Euler kernel $(z-x)^{-\nu}$.
These series expansions are used to obtain azimuthal Fourier and
hyperspherical harmonic expansions for a fundamental solution of the
polyharmonic equation on~$\R^d$.

\begin{theorem}\label{geneneralizationofgeneratingfuncitonforgegenbauerpolyJAC}
Let
$\alpha,\beta>-1$,
such that if $\alpha,\beta\in(-1,0)$ then $\alpha+\beta+1\ne 0$,
$x,z,\nu\in\C$, with
$z\in\C\setminus(-\infty,1]$ on any ellipse with foci at $\pm 1$ and
$x$ in the interior of that ellipse. Then
\begin{gather}
 \frac{1}{(z-x)^\nu}=
\frac{(z-1)^{\alpha+1-\nu}(z+1)^{\beta+1-\nu}}
{2^{\alpha+\beta+1-\nu}}
\nonumber\\
 \hphantom{\frac{1}{(z-x)^\nu}=}{} \times
\sum_{n=0}^\infty
\frac{(\alpha+\beta+2n+1)\Gamma(\alpha+\beta+n+1)(\nu)_n}
{\Gamma(\alpha+1+n)\Gamma(\beta+1+n)}
Q_{n+\nu-1}^{(\alpha+1-\nu,\beta+1-\nu)}(z)
P_n^{(\alpha,\beta)}(x).\!\!\!\!
\label{biggeneralizationgegen2JAC}
\end{gather}
\end{theorem}

\begin{note}\label{NoteAdded}
It has been brought to the author's attention by Tom Koornwinder that
Theorem~\ref{geneneralizationofgeneratingfuncitonforgegenbauerpolyJAC} for $\nu=-n$,
$n\in\N_0$, specializes to formula (21) in~\cite{Koekoek1999}, namely
\begin{gather*}
(z-x)^n=(-2)^nn! \Gamma(\alpha+\beta+1)
\sum_{k=0}^n \frac{(\alpha+\beta+2k+1)(\alpha+\beta+1)_k}
{\Gamma(\alpha+\beta+n+k+2)}\\
\hphantom{(z-x)^n=}{}\times
P_{n-k}^{(-\alpha-n-1,-\beta-n-1)}(z)
P_k^{(\alpha,\beta)}(x).
\end{gather*}
This equivalence is provided by the interesting identity
\begin{gather*}
P_{n-k}^{(-\alpha-n-1,-\beta-n-1)}(z)=
\frac{(-1)^{n+k}\Gamma(\alpha+\beta+n+k+2)(z-1)^{\alpha+n+1}(z+1)^{\beta+n+1}}
{2^{\alpha+\beta+2n+1}(n-k)!\,\Gamma(\alpha+k+1)\Gamma(\beta+k+1)}\\
\hphantom{P_{n-k}^{(-\alpha-n-1,-\beta-n-1)}(z)=}{}\times
Q_{k-n-1}^{(\alpha+n+1,\beta+n+1)}(z),
\end{gather*}
which can be obtained by comparison of Gauss hypergeometric
representations.
\end{note}

\begin{proof}\sloppy Consider the generating function for Gegenbauer polynomials
(see, e.g., \cite[(18.12.4)]{NIST})
\begin{gather}
\frac{1}{(1+\rho^2-2\rho x)^\nu}=\sum_{n=0}^\infty \rho^nC_n^\nu(x)
=\sum_{n=0}^\infty\rho^n\frac{(2\nu)_n}{(\nu+\frac12)_n}P_n^{(\nu-1/2,\nu-1/2)}(x),
\label{geggenfun}
\end{gather}
where we have expressed the Gegenbauer polynomial as a symmetric
Jacobi polynomial \mbox{using~(\ref{gegenjac})}.
Utilizing~(\ref{connectionJac}) in~(\ref{geggenfun}), reversing the order of the summations, and shifting the~$n$
index yields
\begin{gather*}
 \frac{1}{(1+\rho^2-2\rho x)^\nu}=
\sum_{k=0}^\infty
\frac{\rho^k \Gamma(\alpha+\beta+k+1)}
{\Gamma(\alpha+\beta+2k+1)}P_k^{(\alpha,\beta)}(x)\\
\hphantom{\frac{1}{(1+\rho^2-2\rho x)^\nu}=}{}
\times
\sum_{n=0}^\infty
\frac{\rho^n(2\nu)_{n+k}(\nu+k+\frac12)_n(2\nu+n+k)_k}
{n!(\nu+\frac12)_{n+k}}\\
\hphantom{\frac{1}{(1+\rho^2-2\rho x)^\nu}=}{}
\times
\,{}_3F_2\left(
\begin{matrix}
-n,n+2\nu+2k,\alpha+k+1\\
\nu+k+\frac12,\alpha+\beta+2k+2
\end{matrix}
;1
\right).
\end{gather*}
Taking advantage of standard properties such as
(\ref{pochgamma}), (\ref{pochanpk}) produces
\begin{gather*}
 \frac{1}{(1+\rho^2-2\rho x)^\nu}=
\frac{\sqrt{\pi}}{2^{2\nu-1}\Gamma(\nu)}
\sum_{k=0}^\infty
\frac{\rho^k \Gamma(2\nu+2k)\Gamma(\alpha+\beta+k+1)}
{\Gamma(\nu+k+\frac12)\Gamma(\alpha+\beta+2k+1)}P_k^{(\alpha,\beta)}(x)\\
\hphantom{\frac{1}{(1+\rho^2-2\rho x)^\nu}=}{}
\times\sum_{n=0}^\infty\frac{\rho^n(2\nu+2k)_n}{n!}
\,{}_3F_2\left(
\begin{matrix}
-n,n+2\nu+2k,\alpha+k+1\\
\nu+k+\frac12,\alpha+\beta+2k+2
\end{matrix}
;1
\right).
\end{gather*}
We substitute the def\/inition of the ${}_3F_2$ generalized
hypergeometric function
(cf.~(\ref{defnpFq}))
in the sum over $n$, and as previously, reverse the order of the two
summations and shift the summation index.  It then follows using the
duplication formula and
(\ref{pochgamma})--(\ref{binomial}),
that one has
\begin{gather}
 \sum_{n=0}^\infty\frac{\rho^n(2\nu+2k)_n}{n!}
\,{}_3F_2\left(
\begin{matrix}
-n,n+2\nu+2k,\alpha+k+1\\
\nu+k+\frac12,\alpha+\beta+2k+2
\end{matrix}
;1
\right)\nonumber\\
\qquad{}
=\frac{1}{(1-\rho)^{2\nu+2k}}
\,{}_2F_1\left(
\begin{matrix}
\nu+k,\alpha+k+1\\
\alpha+\beta+2k+2
\end{matrix}
;\frac{-4\rho}{(1-\rho)^2}
\right).
\label{3F22F1}
\end{gather}
If we apply the right-hand side of~(\ref{3F22F1})
to~(\ref{defnJacobifnsecondkind})
noting Theorem~12.7.3 (expansion of an analytic function in terms of
orthogonal polynomials) in~\cite{Szego} to obtain the regions of convergence,
we obtain the desired result.
\end{proof}

\begin{corollary}\label{geneneralizationofgeneratingfuncitonforgegenbauerpoly}
Let $\nu\in\C\setminus-\N_0$, with $\mu\in(-\frac12,\infty)\setminus\{0\}$,
and $z\in\C\setminus(-\infty,1]$ on any ellipse with foci at $\pm 1$
with $x$ in the interior of that ellipse. Then
\begin{gather}
\frac
{1}{(z-x)^\nu}
=\frac{2^{\mu+1/2}\Gamma(\mu)e^{i\pi(\mu-\nu+1/2)}}{\sqrt{\pi}\,\Gamma(\nu)
{(z^2-1)^{(\nu-\mu)/2-1/4}}}
\sum_{{n}=0}^\infty({n}+\mu)
Q_{{n}+\mu-1/2}^{\nu-\mu-1/2}(z)
C_{n}^{\mu}(x).
\label{biggeneralizationgegen2GEG}
\end{gather}
\end{corollary}

\begin{proof}Let $\alpha=\beta=\mu-\frac12$ in Theorem~\ref{geneneralizationofgeneratingfuncitonforgegenbauerpolyJAC},
and use~(\ref{QJACQLEG}) and the def\/inition of the Gegenbauer polynomial
in terms of a symmetric Jacobi polynomial~(\ref{gegenjac}).
The points $\nu\in-\N_0$ which must be removed are singularities
originating from the associated Legendre function of the second kind on
the right-hand side of (\ref{biggeneralizationgegen2GEG}).
This complete the proof.
\end{proof}

Note that these singularities are removable and correspond to
non-negative integer powers of the binomial $z-x$. See Note~\ref{NoteAdded}. These singularities can be removed
by taking the limits as~$\nu$ approaches them.

\begin{corollary}\label{MYCOROLLARY}
Let $d\ge 3$, $\nu\in\C\setminus\{0,2,4,\ldots\}$,
$\bfx,\bfxp\in\R^d$ with $r=\|\bfx\|$,
$r^\prime=\|\bfxp\|$, and $\cos\gamma=(\bfx,\bfxp)/(rr^\prime)$. Then
\begin{gather}
 \|\bfx-\bfxp\|^\nu=
\frac{e^{i\pi(\nu+d-1)/2}
\Gamma\left(\frac{d-2}{2}\right)}
{2\sqrt{\pi}\, \Gamma\left(-\frac{\nu}{2}\right)}
\frac{\left(r_>^2-r_<^2\right)^{(\nu+d-1)/2}}
{\left(rr^\prime\right)^{(d-1)/2}}
\nonumber\\
\hphantom{\|\bfx-\bfxp\|^\nu=}{} \times\sum_{{n}=0}^\infty
\left(2{n}+d-2\right)
Q_{{n}+(d-3)/2}^{(1-\nu-d)/2}
\left(\frac{r^2+{r^\prime}^2}{2rr^\prime}\right)
C_{n}^{d/2-1}(\cos\gamma),
\label{expandgegenpowq}
\end{gather}
where $r_\lessgtr={\min \atop \max}\{r,r^\prime\}$.
\end{corollary}

\begin{proof}Map $\nu\mapsto-\nu/2$
in (\ref{biggeneralizationgegen2GEG})
and substitute
$\|\bfx-\bfxp\|=\sqrt{2rr'}\sqrt{z-x}$, $\bfx,\bfxp\in\R^d$ for $d\ge 3$
(\ref{distsph})
with $z=(r^2+{r'}^2)/(2rr')$, $x=\cos\gamma$, i.e.,
\[
\frac{1}{(z-x)^\nu}\mapsto (\sqrt{z-x})^{\nu}
=\frac{1}{(2rr')^{\nu/2}}\|\bfx-\bfxp\|^\nu.
\]
This maps the singularities on the right-hand side
of (\ref{biggeneralizationgegen2GEG})
at $\nu\in-\N_0$ to singularities at $\nu=0,2,4,\ldots$.
\end{proof}

\begin{corollary}\label{geneneralizationofgeneratingfuncitonforgegenbauerpolyFOURIER}
Let $\nu\in\C\setminus-\N_0$,
and $x,z\in\C$ such that $z\in\C\setminus(-\infty,1]$ lie on
any ellipse with foci at $\pm 1$ with  $x$ in the interior
of that ellipse. Then
\begin{gather}
\frac
{1}{(z-x)^\nu}
=\frac{\sqrt{2}\,e^{-i\pi(\nu-1/2)}
}{\sqrt{\pi}\,\Gamma(\nu)
(z^2-1)^{\nu/2-1/4}
}
\sum_{{n}=0}^\infty \epsilon_n T_n(x)
Q_{n-1/2}^{\nu-1/2}(z).
\label{biggeneralizationgegen2FOURIER}
\end{gather}
\end{corollary}

\begin{proof}
Take the limit as $\mu\to 0$ on the right-hand side of
(\ref{biggeneralizationgegen2GEG})
and use
(\ref{limitcheby}).
\end{proof}

 Note that
(\ref{biggeneralizationgegen2FOURIER}) is given in
\cite[(3.10)]{CohlDominici}, so~(\ref{biggeneralizationgegen2JAC})
and~(\ref{biggeneralizationgegen2GEG})
represent generalizations of
that formula.
\begin{corollary}\label{MYOTHERCOROLLARY}
Let $d\ge 2$, $\nu\in\C\setminus\{0,2,4,\ldots\}$,
$\bfx,\bfxp\in\R^d$.
Then
\begin{gather*}
 \|\bfx-\bfxp\|^\nu=
\frac{\sqrt{2}e^{i\pi(\nu+1)/2}(2RR')^{\nu/2}}
{\sqrt{2}\,\Gamma\left(-\frac{\nu}{2}\right)(\chi^2-1)^{-(\nu+1)/4}}
\sum_{m=0}^\infty\epsilon_m T_m(\cos(\phi-\phi'))Q_{m-1/2}^{-(\nu+1)/2}(\chi),
\end{gather*}
where the toroidal parameter $\chi\in[1,\infty)$ is defined in~\eqref{toroidalparameter}  such that $\chi=1$ when $\bfx=\bfxp$.
\end{corollary}

\begin{proof}
Combine
(\ref{algebraicdist}) and Corollary~\ref{geneneralizationofgeneratingfuncitonforgegenbauerpolyFOURIER}.
\end{proof}

\section{Addition theorems in Vilenkin's polyspherical coordinates}
\label{Multisummationadditiontheorems}

In this section we construct power-law addition theorems
in Vilenkin's polyspherical coordinates.

\subsection[Power-law addition theorem on $\R^{d}$ for $d\ge 3$
in standard polyspherical coordinates]{Power-law addition theorem on $\boldsymbol{\R^{d}}$ for $\boldsymbol{d\ge 3}$\\
in standard polyspherical coordinates}

In standard
polyspherical coordinates (\ref{standardsph}) we have the
following multi-summation power-law addition theorem.

\begin{theorem}\label{additiontheoremrddgt3standard}
Let $\nu\in\C\setminus\{2m,2m+2,2m+4,\ldots\}$, $m\in\Z$, $\theta_i\in[0,\pi]$, for $1\le i \le d-2$, $r,r'\in[0,\infty)$ with
$d\ge 3$. Then
\begin{gather}
 Q_{m-1/2}^{-(\nu+1)/2}(\chi)=\sqrt{2}e^{i\pi(d-2)/2}\pi^{(d-4)/2}
\left(2rr^\prime\prod_{i=1}^{d-2}\sin\theta_i{\sin\theta_i'}\right)^{-\nu/2}\nonumber\\
\hphantom{Q_{m-1/2}^{-(\nu+1)/2}(\chi)=}{} \times\left(\chi^2-1\right)^{-(\nu+1)/4}
\frac{
\left(r_>^2-r_<^2\right)^{(\nu+d-1)/2}}
{\left(rr^\prime\right)^{(d-1)/2}}\nonumber\\
\hphantom{Q_{m-1/2}^{-(\nu+1)/2}(\chi)=}{}\times\sum_{l_{d-2}=m}^\infty\frac{(2l_{d-2}+1)(l_{d-2}-m)!}{(l_{d-2}+m)!}
{\rm P}_{l_{d-2}}^m\left(\cos\theta_{d-2}\right)
{\rm P}_{l_{d-2}}^m\left(\cos\theta_{d-2}^\prime\right)\nonumber\\
\hphantom{Q_{m-1/2}^{-(\nu+1)/2}(\chi)=}{} \times\sum_{l_{d-3}=l_{d-2}}^\infty
\Theta_{d-3}^d\left(l_{d-3},l_{d-2};\theta_{d-3}\right)
\Theta_{d-3}^d\left(l_{d-3},l_{d-2};\theta_{d-3}^\prime\right)\cdots\nonumber\\
\hphantom{Q_{m-1/2}^{-(\nu+1)/2}(\chi)=}{}\times\sum_{{l_2}=l_3}^\infty
\Theta_{2}^d\left(l_2,l_3;\theta_2\right)
\Theta_{2}^d\left(l_2,l_3;\theta_2^\prime\right)\nonumber\\
\hphantom{Q_{m-1/2}^{-(\nu+1)/2}(\chi)=}{} \times\sum_{{l_1}=l_2}^\infty
\Theta_{1}^d\left(l_1,l_2;\theta_1\right)
\Theta_{1}^d\left(l_1,l_2;\theta_1^\prime\right)
Q_{l_1+(d-3)/2}^{(1-\nu-d)/2}
\left(\frac{r^2+{r^\prime}^2}{2rr^\prime}\right),
\label{bigstandardpolythm}
\end{gather}
where
\begin{gather*}
 \chi=
\frac
{r^2+{r^\prime}^2-2rr^\prime
{ \sum\limits_{i=1}^{d-2}\cos\theta_i{\cos\theta_i'}
\prod\limits_{j=1}^{i-1}\sin\theta_j\sin\theta_j'}}
{  2rr^\prime
\prod\limits_{i=1}^{d-2}\sin\theta_i{\sin\theta_i'}
}.
\end{gather*}
\end{theorem}

\begin{proof}
If we adopt standard polyspherical coordinates (\ref{standardsph})
(see Fig.~\ref{Fig:stansph}),
then one can obtain an eigenfunction expansion for a power-law fundamental
solution of the polyharmonic equation in standard polyspherical coordinates
using the Gegenbauer expansion (\ref{expandgegenpowq}) with the
addition theorem for hyperspherical harmonics (\ref{addthmhypsph}),
and the normalized standard hyperspherical harmonics
(\ref{standardhypersphericalharmonic}), obtaining
\begin{gather}
 \|\bfx-\bfxp\|^\nu=
\frac{4e^{i\pi(\nu+d-1)/2}\pi^{(d-1)/2}}
{\Gamma\left(-\frac{\nu}{2}\right)}
\frac{\left(r_>^2-r_<^2\right)^{(\nu+d-1)/2}}
{\left(rr^\prime\right)^{(d-1)/2}}\nonumber\\
\hphantom{\|\bfx-\bfxp\|^\nu=}{}
\times\sum_{l_1,K}
Q_{l_1+(d-3)/2}^{(1-\nu-d)/2}
\left(\frac{r^2+{r^\prime}^2}{2rr^\prime}\right)
Y_{l_1}^K(\widehat {\mathbf{x}})\overline{Y_{l_1}^K
(\widehat{{\mathbf{x}}}^\prime)}.
\label{powgegenexpandedwithq}
\end{gather}
If we expand the product of polyspherical harmonics
in (\ref{powgegenexpandedwithq})
with (\ref{standardhypersphericalharmonic}) after reversing the order of the summations,
we obtain
\begin{gather}
\|\bfx-\bfxp\|^\nu=\sum_{m=0}^\infty\cos(m(\phi-\phi^\prime))
\frac{\pi^{(d-3)/2}e^{i\pi(\nu+d-1)/2}\epsilon_m}{\Gamma\left(-\frac{\nu}{2}\right)}
\frac{\left(r_>^2-r_<^2\right)^{(\nu+d-1)/2}}
{\left(rr^\prime\right)^{(d-1)/2}}\nonumber\\
\hphantom{\|\bfx-\bfxp\|^\nu=}{}\times\sum_{l_{d-2}=m}^\infty\frac{(2l_{d-2}+1)(l_{d-2}-m)!}{(l_{d-2}+m)!}
{\rm P}_{l_{d-2}}^m\left(\cos\theta_{d-2}\right)
{\rm P}_{l_{d-2}}^m\left(\cos\theta_{d-2}^\prime\right)\nonumber\\
\hphantom{\|\bfx-\bfxp\|^\nu=}{}\times\sum_{l_{d-3}=l_{d-2}}^\infty
\Theta_{d-3}^d\left(l_{d-3},l_{d-2};\theta_{d-3}\right)
\Theta_{d-3}^d\left(l_{d-3},l_{d-2};\theta_{d-3}^\prime\right)\cdots \nonumber\\
\hphantom{\|\bfx-\bfxp\|^\nu=}{}\times\sum_{{l_2}=l_3}^\infty
\Theta_{2}^d\left(l_2,l_3;\theta_2\right)
\Theta_{2}^d\left(l_2,l_3;\theta_2^\prime\right)\nonumber\\
\hphantom{\|\bfx-\bfxp\|^\nu=}{}\times\sum_{{l_1}=l_2}^\infty
\Theta_{1}^d\left(l_1,l_2;\theta_1\right)
\Theta_{1}^d\left(l_1,l_2;\theta_1^\prime\right)
Q_{l_1+(d-3)/2}^{(1-\nu-d)/2}
\left(\frac{r^2+{r^\prime}^2}{2rr^\prime}\right),
\label{expandedhypsphstandardwithq}
\end{gather}
where $\Theta_j^d(l_j,l_{j+1};\theta)$, for $1\le j \le d-2$ is def\/ined
in~(\ref{stansphharmoncis}).

The Fourier expansion for a power-law fundamental solution of the
polyharmonic equation in standard polyspherical coordinates is obtained
by substituting the expansion in terms of Chebyshev polynomials
of the f\/irst kind~(\ref{biggeneralizationgegen2FOURIER}) in
the algebraic expression for a power-law fundamental solution of
the polyharmonic equation (cf.~(\ref{algebraicdist})).  This results in
\begin{gather}
 \|\bfx-\bfxp\|^\nu=\sqrt{\frac{\pi}{2}}
\frac{e^{i\pi(\nu+1)/2}}{\Gamma\left(-\nu/2\right)}
\left(2rr^\prime
\prod_{i=1}^{d-2}\sin\theta_i{\sin\theta_i'}
\right)^{\nu/2}\left(\chi^2-1\right)^{(\nu+1)/4}
\nonumber\\
 \hphantom{\|\bfx-\bfxp\|^\nu=}{}
 \times\sum_{m=-\infty}^\infty e^{im(\phi-\phi^\prime)}
Q_{m-1/2}^{-(\nu+1)/2}(\chi).
\label{fourierexpansionpowerstandsph}
\end{gather}
By comparing the Fourier coef\/f\/icients of
(\ref{expandedhypsphstandardwithq})
with~(\ref{fourierexpansionpowerstandsph}), we complete the proof of
this \linebreak theorem.
\end{proof}

This is just one example of a derived multi-summation addition theorem for arbitrary
dimensions.  There are an unlimited number of such straightforward examples to generate.
In the next section we derive another example which is valid on~$\R^d$
where $d$ is given by a power of two, generalized Hopf coordinates~(\ref{genHopf}).

\subsection[Power-law addition theorem on $\R^{2^q}$ for $q\ge 2$ in generalized Hopf coordinates]{Power-law addition theorem on $\boldsymbol{\R^{2^q}}$ for $\boldsymbol{q\ge 2}$ \\ in generalized Hopf coordinates}

In generalized Hopf coordinates (\ref{genHopf}) we have the
following multi-summation power-law addition theorem.

\begin{theorem}\label{additiontheoremr2q}
Let $\nu\in\C\setminus\{2m,2m+2,2m+4,\ldots\}$, $m_1\in\Z$, $r,r'\in[0,\infty)$,
$\vartheta_i\in\big[0,\frac{\pi}{2}\big]$ with $1\le i \le 2^{q-1}-1$,
$\phi_i\in[0,2\pi)$ such that $1\le i \le 2^{q-1}$. Then
\begin{gather}
 Q_{m_1-1/2}^{-(\nu+1)/2}(\chi)=\frac{ -\left(\prod\limits_{j=1}^{q-1}
\cos\vartheta_{2^{j-1}}
\cos\vartheta_{2^{j-1}}'\right)^{-\nu/2}}{2^{(\nu+1)/2}(\chi^2-1)^{(\nu+1)/4}}
\left(\frac{r_>^2-r_<^2}{rr'}\right)^{(\nu+2^q-1)/2}\nonumber\\
\hphantom{Q_{m_1-1/2}^{-(\nu+1)/2}(\chi)=}{}
 \times\sum_{m_{2}=0}^\infty \epsilon_{m_{2}}\cos(m_{2}(\phi_{2}-\phi_{2}'))
\cdots
\sum_{m_{2^{q-1}}=0}^\infty \epsilon_{m_{2^{q-1}}}\cos(m_{2^{q-1}}
(\phi_{2^{q-1}}-\phi_{2^{q-1}}'))\nonumber\\
\hphantom{Q_{m_1-1/2}^{-(\nu+1)/2}(\chi)=}{}
\times \sum_{n_{2^{q-1}-1}=0}^\infty
\Upsilon_{2^{q-1}-1}^{q}
\left(
\begin{matrix}
n_{2^{q-1}-1}\\
|m_{2^{q-1}-1}|,|m_{2^{q-1}}|
\end{matrix};\vartheta_{2^{q-1}-1} \right)\nonumber\\
\hphantom{Q_{m_1-1/2}^{-(\nu+1)/2}(\chi)=}{}
\times \Upsilon_{2^{q-1}-1}^{q}
\left(
\begin{matrix}
n_{2^{q-1}-1}\\[1pt]
|m_{2^{q-1}-1}|,|m_{2^{q-1}}|
\end{matrix};\vartheta_{2^{q-1}-1}' \right) \cdots\nonumber\\
\hphantom{Q_{m_1-1/2}^{-(\nu+1)/2}(\chi)=}{}
\times \sum_{n_{2^{q-2}}=0}^\infty
\Upsilon_{2^{q-2}}^{q}
\left(
\begin{matrix}
n_{2^{q-2}}\\
|m_{1}|,|m_{2}|
\end{matrix};\vartheta_{2^{q-2}} \right)
\Upsilon_{2^{q-2}}^{q}
\left(
\begin{matrix}
n_{2^{q-2}}\\
|m_{1}|,|m_{2}|
\end{matrix};\vartheta_{2^{q-2}}' \right)\nonumber\\
\hphantom{Q_{m_1-1/2}^{-(\nu+1)/2}(\chi)=}{}
\times \sum_{n_{2^{q-2}-1}=0}^\infty
\Upsilon_{2^{q-2}-1}^{q}
\left(
\begin{matrix}
n_{2^{q-2}-1}\\
l_{2^{q-1}-2},l_{2^{q-1}-1}
\end{matrix};\vartheta_{2^{q-2}-1} \right)\nonumber\\
\hphantom{Q_{m_1-1/2}^{-(\nu+1)/2}(\chi)=}{}
\times
\Upsilon_{2^{q-2}-1}^{q}
\left(
\begin{matrix}
n_{2^{q-2}-1}\\
l_{2^{q-1}-2},l_{2^{q-1}-1}
\end{matrix};\vartheta_{2^{q-2}-1}' \right)
\cdots\nonumber\\
\hphantom{Q_{m_1-1/2}^{-(\nu+1)/2}(\chi)=}{}
\times \sum_{n_{1}=0}^\infty
\Upsilon_{1}^{q}
\left(
\begin{matrix}
n_{1}\\
l_{2},l_{3}
\end{matrix};\vartheta_{1} \right)
\Upsilon_{1}^{q}
\left(
\begin{matrix}
n_{1}\\
l_{2},l_{3}
\end{matrix};\vartheta_{1}' \right)
Q_{2n_{1}+l_{2}+l_{3}+(2^{q}-3)/2}^{(1-\nu-2^q)/2}
\left(\frac{r^2+{r'}^2}{2rr'}\right),\!\!\!
\label{Upsilonaddthm}
\end{gather}
where
\begin{gather}
\chi=
\frac{r^2+{r^\prime}^2-2rr'\cos\gamma+2rr^\prime\cos(\phi_1-\phi_1')\,
{  \prod\limits_{j=1}^{q-1}\cos\vartheta_{2^{j-1}}\cos\vartheta_{2^{j-1}}'}}
{{  2rr^\prime
\prod\limits_{j=1}^{q-1}\cos\vartheta_{2^{j-1}}{\cos\vartheta_{2^{j-1}}'}}
}.
\label{chiHopf}
\end{gather}
\end{theorem}

\begin{proof}
If we adopt generalized Hopf coordinates (\ref{genHopf})
(see Fig.~\ref{Fig:genHopf}),
then we can use the corresponding harmonics~(\ref{genHopfcoordinatesthis})
in combination with the addition theorem for
hyperspherical harmonics~(\ref{addthmhypsph}).
We compare the Gegenbauer expansion
for powers of the distance~(\ref{expandgegenpowq}) with the Fourier expansion
\begin{gather}
\|\bfx-\bfxp\|^\nu=\frac
{\sqrt{2}ie^{i\pi\nu/2}{\left(2rr'\prod\limits_{j=1}^{q-1}
\cos\vartheta_{2^{j-1}}\cos\vartheta_{2^{j-1}}'\right)^{\nu/2}}}
{\sqrt{\pi}\,\Gamma\left(\frac{-\nu}{2}\right)
(\chi^2-1)^{-(\nu+1)/4}}\nonumber\\
\hphantom{\|\bfx-\bfxp\|^\nu=}{}
\times
\sum_{m_1=-\infty}^\infty e^{im_1(\phi_1-\phi_1^\prime)}
Q_{m_1-1/2}^{-(\nu+1)/2}(\chi),
\label{fourierexpansionpowerstandsphHopf}
\end{gather}
where $\chi>1$ is given by (\ref{chiHopf}).
Notice that $\chi$ is independent of $\phi_1-\phi_1'$.
By using the Gegenbauer expansion~(\ref{expandgegenpowq}) and inserting the appropriate Gegenbauer
polynomial using the addition theorem for hyperspherical harmonics~(\ref{addthmhypsph}), we obtain
\begin{gather}
 \|\bfx-\bfxp\|^\nu=
\frac{-2i\pi^{2^{q-1}}e^{i\pi\nu/2}(r_>^2-r_<^2)^{(\nu+2^q-1)/2}}
{\sqrt{\pi}\,\Gamma\left(\frac{-\nu}{2}\right)(rr')^{(2^q-1)/2}}\nonumber\\
\hphantom{\|\bfx-\bfxp\|^\nu=}{}
 \times\sum_{l_1,K}
Q_{l_1+(2^q-3)/2}^{(1-\nu-2^q)/2}
\left(\frac{r^2+{r^\prime}^2}{2rr^\prime}\right)
Y_{l_1}^K( \widehat {\mathbf{x}} )\overline{Y_{l_1}^K
( \widehat{\mathbf{x}}^\prime)}.
\label{powgegenexpandedwithqHopf}
\end{gather}
By expanding the product of polyspherical harmonics
in~(\ref{powgegenexpandedwithqHopf})
with~(\ref{genHopfcoordinatesthis}) expressed in terms of surrogate
quantum numbers and reversing the order of the sums, we obtain
a multi-summation expression
for the power of the Euclidean distance between two points in generalized
Hopf coordinates.  Through comparison of the resulting equation with
the $m_1$ Fourier coef\/f\/icients of~(\ref{fourierexpansionpowerstandsphHopf}),
we derive~(\ref{Upsilonaddthm}),
a multi-summation addition theorem for the associated
Legendre function of the second kind with argument~$\chi$
(cf.~(\ref{chiHopf})).
\end{proof}

\subsection[Power-law addition theorems on $\R^3$]{Power-law addition theorems on $\boldsymbol{\R^3}$}

In $d=3$ there are two ways to construct polyspherical coordinates, with
trees of type ${\mathbf{b}}^\prime{\mathbf{a}}$
(see Fig.~\ref{Fig:tree23}b)
and $\mathbf{ba}$
(see Fig.~\ref{Fig:tree23}c).   We only treat the f\/irst tree
since the addition theorem from the second tree is trivially obtained
from the f\/irst.

\subsubsection[Type $ba$ coordinates]{Type $\mathbf{ba}$ coordinates}

\begin{corollary}\label{additiontheoremba}
Let $\nu\in\C\setminus\{2m,2m+2,2m+4,\ldots\}$, $m\in\N_0$, $\theta,\theta'\in[0,\pi]$, $r,r'\in [0,\infty)$.
Then
\begin{gather}
 Q_{m-1/2}^{-(\nu+1)/2}(\chi)=i\sqrt{\pi} 2^{-(\nu+3)/2}(\sin\theta\sin\theta^\prime)^{-\nu/2}(\chi^2-1)^{-(\nu+1)/4}
\left(\frac{r_>^2-r_<^2}{rr^\prime}\right)^{(\nu+2)/2}\nonumber\\
\hphantom{Q_{m-1/2}^{-(\nu+1)/2}(\chi)=}{}
\times\sum_{l=m}^\infty (2l+1)\frac{(l-m)!}{(l+m)!}Q_l^{-(\nu+2)/2}\left(\frac{r^2+{r^\prime}^2}{2rr^\prime} \right)
{\rm P}_l^m(\cos\theta) {\rm P}_l^m(\cos\theta^\prime),
\label{addthm3ba}
\end{gather}
where
\begin{gather}
\chi=\frac{r^2+{r^\prime}^2-2rr^\prime\cos\theta\cos\theta^\prime}
{2rr^\prime\sin\theta\sin\theta^\prime}.
\label{chiba}
\end{gather}
\end{corollary}

\begin{proof}\looseness=1
Taking $d=3$ in (\ref{bigstandardpolythm}) for
type $\mathbf{ba}$ coordinates converts~$\chi$ to~(\ref{chiba}) and the relevant Gegenbauer polynomials reduce to
Ferrers functions through~(\ref{GegenLeg}).
This completes the proof.
\end{proof}

  Equation (\ref{addthm3ba}) is a generalization of one of the main results
of~\cite{CRTB}. This can be observed if you substitute
$\nu=-1$ in (\ref{addthm3ba}) (this corresponds to a fundamental solution of Laplace's
equation on $\R^3$),
then the associated Legendre function of the second kind on the right-hand side reduces
to an elementary function through \cite[(8.6.11)]{Abra}, producing
\begin{gather*}
 Q_{m-1/2}(\chi)=\pi\sqrt{\sin\theta\sin\theta^\prime}\sum_{l=|m|}^\infty \frac{(l-m)!}{(l+m)!}
\left(\frac{r_<}{r_>} \right)^{l+1/2}{\rm P}_l^m(\cos\theta) {\rm P}_l^m(\cos\theta^\prime).
\end{gather*}

\subsection[Power-law addition theorems on $\R^4$]{Power-law addition theorems on $\boldsymbol{\R^4}$}

In $d=4$ there are f\/ive ways to construct polyspherical coordinates, with
trees of type
${\mathbf{b}}^2{\mathbf{a}}$
(see Fig.~\ref{Fig:tree4}a),
${\mathbf{bb}}^\prime {\mathbf{a}}$
(see Fig.~\ref{Fig:tree4}b),
${\mathbf{b}}^\prime{\mathbf{ba}}$
(see Fig.~\ref{Fig:tree4}c),
${\mathbf{b}}^{\prime 2}{\mathbf{a}}$
(see Fig.~\ref{Fig:tree4}d),
${\mathbf{ca}}^2$
(see Fig.~\ref{Fig:tree4}e).
We only treat the f\/irst and the f\/ifth trees
since the addition theorems from the second, third and fourth trees
are trivially obtained from the f\/irst.

\subsubsection[Type $b^2a$ coordinates]{Type $\boldsymbol{{\mathbf{b}}^2{\mathbf{a}}}$ coordinates}

\begin{corollary}\label{additiontheoremb2a}
Let $\nu\in\C\setminus\{2m,2m+2,2m+4,\ldots\}$, $m\in\Z$, $r,r'\in[0,\infty)$,
$\theta_1,\theta_1',\theta_2,\theta_2'\in[0,\pi]$.  Then
\begin{gather}
   Q_{m-1/2}^{-(\nu+1)/2}(\chi)=
\frac{-1}{2^{(\nu+1)/2}}
\left(\frac{r_>^2-r_<^2}{rr^\prime} \right)^{(\nu+3)/2}\left(\chi^2-1\right)^{-(\nu+1)/4}
(\sin\theta_1\sin\theta_1^\prime\sin\theta_2\sin\theta_2^\prime)^{-\nu/2}\nonumber\\
\hphantom{Q_{m-1/2}^{-(\nu+1)/2}(\chi)=}{}
\times\sum_{l_2=|m|}^\infty \frac{2^{2l_2}(2l_2+1)(l_2!)^2(l_2-m)!}{(l_2+m)!}(\sin\theta_1\sin\theta_1^\prime)^{l_2}
{\rm P}_{l_2}^m(\cos\theta_2) {\rm P}_{l_2}^m(\cos\theta_2^\prime)\nonumber\\
\hphantom{Q_{m-1/2}^{-(\nu+1)/2}(\chi)=}{}
\times\sum_{l_1=l_2}^\infty\frac{(l_1+1)(l_1-l_2)!}{(l_1+l_2+1)!}Q_{l_1+1/2}^{-(\nu+3)/2}
\left(\frac{r^2+{r^\prime}^2}{2rr^\prime} \right)\nonumber\\
\hphantom{Q_{m-1/2}^{-(\nu+1)/2}(\chi)=}{}
\times
C_{l_1-l_2}^{l_2+1}(\cos\theta_1)
C_{l_1-l_2}^{l_2+1}(\cos\theta_1^\prime),
\label{addthm4b2a}
\end{gather}
where
\begin{gather*}
\chi=\frac{r^2+{r^\prime}^2-2rr^\prime\cos\theta_1\cos\theta_1^\prime
-2rr^\prime\sin\theta_1\sin\theta_1^\prime\cos\theta_2\cos\theta_2^\prime}
{2rr^\prime\sin\theta_1\sin\theta_1^\prime\sin\theta_2\sin\theta_2^\prime}.
\end{gather*}
\end{corollary}
\begin{proof}Taking $d=4$ in (\ref{bigstandardpolythm})
completes the proof.
\end{proof}

If you substitute $\nu=-2$ (a fundamental solution for the Laplacian on $\R^4$) in (\ref{addthm4b2a})
then the Legendre functions of the second kind reduce to elementary functions through
\cite[(8.6.10-11)]{Abra}, and one obtains the following
\begin{gather*}
  \frac{(\chi^2-1)^{-1/2}}{(\chi+\sqrt{\chi^2-1})^{m}}
=2\sin\theta_1\sin\theta_1^\prime\sin\theta_2\sin\theta_2^\prime \\
\hphantom{\frac{(\chi^2-1)^{-1/2}}{(\chi+\sqrt{\chi^2-1})^{m}}=}{}
\times\sum_{l_2=|m|}^\infty \!\!\frac{2^{2l_2}(2l_2+1)(l_2!)^2(l_2-m)!}{(l_2+m)!}(\sin\theta_1\sin\theta_1^\prime)^{l_2}
{\rm P}_{l_2}^m(\cos\theta_2) {\rm P}_{l_2}^m(\cos\theta_2^\prime) \\
\hphantom{\frac{(\chi^2-1)^{-1/2}}{(\chi+\sqrt{\chi^2-1})^{m}}=}{}
\times\sum_{l_1=l_2}^\infty
\frac{(l_1-l_2)!}{(l_1+l_2+1)!}\left(\frac{r_<}{r_>} \right)^{l_1+1}
C_{l_1-l_2}^{l_2+1}(\cos\theta_1)
C_{l_1-l_2}^{l_2+1}(\cos\theta_1^\prime).
\end{gather*}

\subsubsection[Type $ca^2$ coordinates]{Type $\boldsymbol{{\mathbf{ca}}^2}$ coordinates}

\begin{corollary}\label{additiontheoremca2}
Let $\nu\in\C\setminus\{2m,2m+2,2m+4,\ldots\}$, $m_1\in\Z$, $r,r'\in[0,\infty)$,
$\vartheta,\vartheta'\in[0,\frac{\pi}{2}]$, $\phi_2,\phi_2'\in[0,2\pi)$.  Then
\begin{gather}
 Q_{m_1-1/2}^{-(\nu+1)/2}(\chi)=
\frac{-2^{-(\nu+1)/2}}
{\left(\chi^2-1\right)^{(\nu+1)/4}}
\left( \frac{r_>^2-r_<^2}{rr^\prime} \right)^{(\nu+3)/2}
\left( \cos\vartheta\cos\vartheta^\prime\right)^{|m_1|-\nu/2}\nonumber\\
\hphantom{Q_{m_1-1/2}^{-(\nu+1)/2}(\chi)=}{}
\times\sum_{m_2=0}^\infty \epsilon_{m_2}\cos(m_2(\phi_2-\phi_2^\prime))
(\sin\vartheta\sin\vartheta^\prime)^{m_2}\nonumber\\
\hphantom{Q_{m_1-1/2}^{-(\nu+1)/2}(\chi)=}{}
\times \sum_{n=0}^\infty
\frac{(2n+|m_1|+m_2+1)(|m_1|+m_2+n)!n!}{(|m_1|+n)!(m_2+n)!}\nonumber\\
\hphantom{Q_{m_1-1/2}^{-(\nu+1)/2}(\chi)=}{}
\times Q_{2n+|m_1|+m_2+1/2}^{-(\nu+3)/2}\left(\frac{r^2+{r^\prime}^2}{2rr^\prime} \right)
P_n^{(m_2,|m_1|)}(\cos 2\vartheta)
P_n^{(m_2,|m_1|)}(\cos 2\vartheta^\prime),\!\!\!
\label{addthm4ca2}
\end{gather}
where
\begin{gather*}
\chi=\frac{r^2+{r^\prime}^2-2rr^\prime\sin\vartheta\sin\vartheta^\prime\cos(\phi_2-\phi_2^\prime)}
{2rr^\prime\cos\vartheta\cos\vartheta^\prime}.
\end{gather*}
\end{corollary}

\begin{proof}
Taking $q=2$ in
(\ref{Upsilonaddthm}) completes the proof.
\end{proof}

If you substitute $\nu=-2$ in
(\ref{addthm4ca2}), then the Legendre functions of the second kind reduce to elementary functions through
\cite[(8.6.10-11)]{Abra}, and one obtains the following
\begin{gather}
  \frac{(\chi^2-1)^{-1/2}}{(\chi+\sqrt{\chi^2-1})^{m_1}}
=2\left( \cos\vartheta\cos\vartheta^\prime\right)^{|m_1|+1}
\sum_{m_2=0}^\infty \epsilon_{m_2}\cos\left[m_2(\phi_2-\phi_2^\prime)\right]
(\sin\vartheta\sin\vartheta^\prime)^{|m_2|}
\nonumber\\
\hphantom{\frac{(\chi^2-1)^{-1/2}}{(\chi+\sqrt{\chi^2-1})^{m_1}}=}{}
\times
\sum_{n=0}^\infty
\frac{(|m_1|+|m_2|+n)!n!}{(|m_1|+n)!(|m_2|+n)!}
\left(\frac{r_<}{r_>} \right)^{|m_1|+|m_2|+2n+1}
\nonumber\\
\hphantom{\frac{(\chi^2-1)^{-1/2}}{(\chi+\sqrt{\chi^2-1})^{m_1}}=}{}
\times
 P_n^{(|m_2|,|m_1|)}(\cos 2\vartheta)
P_n^{(|m_2|,|m_1|)}(\cos 2\vartheta^\prime).
\label{addthm4ca2b}
\end{gather}

 Note that in the addition theorems (\ref{addthm4ca2}) and (\ref{addthm4ca2b}), that if you
make the map $\vartheta\mapsto\vartheta-\frac{\pi}{2}$, then
this transformation preserves the addition theorems
such that $m_1\leftrightarrow m_2$.  This transformation is equivalent
to swapping the position of~$\phi_1$ and~$\phi_2$ for the
tree in Fig.~\ref{Fig:tree4}e.

\appendix

\section{Special functions and orthogonal polynomials}
\label{Specialfunctionsandorthogonalpolynomials}

The generalized hypergeometric function
${}_pF_q:\C^p\times(\C\setminus-\N_0)^q\times
\{z\in\C:|z|<1\}\to\C$ can be def\/ined as
\begin{gather}
{}_pF_q\left(
\begin{matrix}
a_1,\ldots,a_p\\
b_1,\ldots,b_q
\end{matrix}
;z
\right):=\sum_{n=0}^\infty
\frac{(a_1)_n\cdots(a_p)_n}{(b_1)_n\cdots(b_q)_n}\frac{z^n}{n!},
\label{defnpFq}
\end{gather}
where the Pochhammer symbol (rising factorial)
$(\cdot)_n:\C\to\C$
for $n\in\N_0$ is def\/ined by
$(z)_n:=\prod\limits_{i=1}^n(z+i-1)$.
Furthermore one has
\begin{gather}
(z)_n=\frac{\Gamma(z+n)}{\Gamma(z)}
\label{pochgamma}
\end{gather}
for $z\in\C\setminus-\N_0$, which implies
\begin{gather}
(z)_{n+k}=(z)_k (z+k)_n,
\label{pochanpk}
\end{gather}
$k\in\N_0$, $z\in\C$.
One also has
\begin{gather}
(-n-k)_k=\frac{(-1)^k(n+k)!}{n!}.
\label{mnmkk}
\end{gather}
In this paper, we will use
two dif\/ferent generalized hypergeometric functions, namely
${}_3F_2$ and the Gauss hypergeometric function ${}_2F_1$
(see for instance Chapter 15 in~\cite{NIST}).
We also use the binomial expansions for $p=1$, $q=0$, namely
\begin{gather}
{}_1F_0\left(\begin{matrix}\alpha\\ -\end{matrix};z\right)=(1-z)^{-\alpha},
\label{binomial}
\end{gather}
where $\alpha,z\in\C$ such that $|z|<1$.
The special functions used in this paper as well as and their properties
can be described in terms of these.

There are many important orthogonal polynomials which can be def\/ined in
terms of a terminating generalized hypergeometric series.  The
Jacobi polynomials $P_n^{(\alpha,\beta)}:\C\to\C$,
for $n\in\N_0$, and $\alpha$, $\beta>-1$
such that if $\alpha,\beta\in(-1,0)$ then $\alpha+\beta+1\ne 0$,
are def\/ined as~\cite[(18.5.7)]{NIST}
\[
P_n^{(\alpha,\beta)}(z):=
\frac{(\alpha+1)_n}{n!}\,
{}_2F_1\left(
\begin{matrix}
-n,n+\alpha+\beta+1\\
\alpha+1
\end{matrix};
\frac{1-z}{2}
\right).
\]
These polynomials are orthogonal with respect to the positive
weight $w:(-1,1)\to[0,\infty)$, $w(x):=(1-x)^\alpha(1+x)^\beta$,
with orthogonality relation
\[
\int_{-1}^1
P_m^{(\alpha,\beta)}(x)
P_n^{(\alpha,\beta)}(x)
(1-x)^\alpha
(1+x)^\beta
dx=\frac{\delta_{m,n}}{\big(N_n^{\alpha,\beta}\big)^2}.
\]
The normalization constant
$N_n^{\alpha,\beta}\in(0,\infty)$
is given by
\[
N_n^{\alpha,\beta}=\sqrt{
\frac
{(2n+\alpha+\beta+1)\Gamma(n+\alpha+\beta+1)n!}
{2^{\alpha+\beta+1}\Gamma(n+\alpha+1)\Gamma(n+\beta+1)}.
}
\]
The connection relation for Jacobi polynomials with two free parameters
is given by
(see for instance~\cite[p.~256]{Ismail})
\begin{gather}
P_n^{(\gamma,\delta)}(x)=\sum_{k=0}^n c_{n,k}(\gamma,\delta;\alpha,\beta) P_k^{(\alpha,\beta)}(x),
\label{connectionJac}
\end{gather}
where $\gamma,\delta>-1$, and
such that if $\gamma,\delta\in(-1,0)$ then $\gamma+\delta+1\ne 0$,
\begin{gather*}
 c_{n,k}(\gamma,\delta;\alpha,\beta)=
\frac{(\gamma+k+1)_{n-k}(n+\gamma+\delta+1)_k\Gamma(\alpha+\beta+k+1)}
{(n-k)!\Gamma(\alpha+\beta+2k+1)} \\
\hphantom{c_{n,k}(\gamma,\delta;\alpha,\beta)=}{}
 \times {}_3F_2\left(
\begin{matrix}
-n+k,n+k+\gamma+\delta+1,\alpha+k+1\\
\gamma+k+1,\alpha+\beta+2k+2
\end{matrix}
;1
\right).
\end{gather*}

Jacobi polynomials with parameters
$\alpha=\beta$ are described as
symmetric and are representable in terms of Gegenbauer polynomials
using (see (6.4.9) in~\cite{AAR})
\begin{gather}
C_n^\nu(x)=
\frac
{(2\nu)_n}
{(\nu+\frac12)_n}
P_n^{(\nu-1/2,\nu-1/2)}(x),
\label{gegenjac}
\end{gather}
where
$\nu\in(-\frac12,\infty)\setminus\{0\}$.
The Chebyshev polynomial of the f\/irst kind $T_{n}:\C\to\C$ is
def\/ined as (see \S~5.7.2 in~\cite{MOS})
\[
T_n(z):=
{}_2F_1\left(\begin{matrix}-n,n\\
\frac12\end{matrix};\frac{1-z}{2}\right),
\]
for $n\in\N_0$.
These can be computed in
terms of Gegenbauer polynomials using
\begin{gather}
\lim_{\mu\to 0}\frac{{n}+\mu}{\mu}C_{n}^\mu(x)=\epsilon_{n} T_{n}(x)
\label{limitcheby}
\end{gather}
(see for instance (6.4.13) in~\cite{AAR}).

The associated Legendre function of the second kind
$Q_\nu^\mu:\C\setminus(-\infty,1]\to\C$, $\nu+\mu\notin-\N$, can be def\/ined
in terms of the Gauss hypergeometric function as follows
\cite[(14.3.7) and \S~14.21]{NIST}
\[
Q_\nu^\mu(z):=\frac{\sqrt{\pi}e^{i\pi\mu}\Gamma(\nu+\mu+1)(z^2-1)^{\mu/2}}
{2^{\nu+1}\Gamma(\nu+\frac32)z^{\nu+\mu+1}}\,
{}_2F_1\left(
\begin{matrix}
\frac{\nu+\mu+1}{2},
\frac{\nu+\mu+2}{2}\\
\nu+\frac32
\end{matrix};
\frac{1}{z^2}
\right),
\]
for $|z|>1$ and elsewhere in $z$ by analytic continuation
of the Gauss hypergeometric function.
One may also def\/ine the associated Legendre function of the second kind
using \cite[entry~24, p.~161]{MOS}
\begin{gather}
 Q_\nu^\mu(z):=
\frac{e^{i\pi\mu}2^\nu\Gamma(\nu+1)\Gamma(\nu+\mu+1)(z+1)^{\mu/2}}
{\Gamma(2\nu+2)(z-1)^{b/2+a+1}}
\,{}_2F_1\left(
\begin{matrix}
\nu+1,\nu+\mu+1\\
2+2\nu
\end{matrix}
;\frac{2}{1-z}
\right),
\label{Qdefntwodivide1mz}
\end{gather}
for $|1-z|>2$.
Similarly, the associated Legendre function of the f\/irst kind can be def\/ined using the
Gauss hypergeometric function  \cite[(14.3.6) and
\S~14.21(i)]{NIST}
\begin{gather*}
P_\nu^\mu(z):=\frac{1}{\Gamma(1-\mu)}\left(\frac{z+1}{z-1}\right)^{\mu/2}
{}_2F_1\left(\begin{matrix}-\nu,\nu+1\\
1-\mu\end{matrix};\frac{1-z}{2}\right),
\end{gather*}
where $|1-z|<2$, and elsewhere in $z$ by analytic continuation.
We can use Whipple's formula to relate the associated Legendre function
of the f\/irst kind with the associated Legendre function of the
second kind.  It is given by
 \cite[(8.2.7)]{Abra}
\begin{gather}
P_{-\mu-1/2}^{-\nu-1/2}
\biggl(\frac{z}{\sqrt{z^2-1}}\biggr)=
\sqrt{\frac{2}{\pi}}
\frac{(z^2-1)^{1/4}e^{-i\mu\pi}}
{\Gamma(\nu+\mu+1)}Q_\nu^\mu(z),
\label{whipple}
\end{gather}
for $\mbox{Re}\,z>0$.
The Ferrers function of the
f\/irst kind (associated Legendre function of the f\/irst kind on-the-cut)
$\mathrm{P}_\nu^\mu:(-1,1)\to\C$ can be def\/ined as
\cite[(14.3.1)]{NIST}
\begin{gather}
\mathrm{P}_\nu^\mu(x):=\frac{1}{\Gamma(1-\mu)}
\left(\frac{1+x}{1-x}\right)^{\mu/2}
{}_2F_1\left(
\begin{matrix}
-\nu,\nu+1\\
1-\mu
\end{matrix};\frac{1-x}{2}\right).
\label{FerrersPdefnGauss2F1}
\end{gather}
There is a relation between certain Gegenbauer polynomials
on $(-1,1)$ and the Ferrers function of the f\/irst kind
(cf.~(8.936.2) in~\cite{Grad}), namely
\begin{gather}
C_{l-m}^{m+1/2}(x)=
\frac{
(-1)^m(1-x^2)^{-m/2}
}
{
(2m-1)!!
}
{\rm P}_l^m(x),
\label{GegenLeg}
\end{gather}
where the double factorial
$\cdot :\{-1,0,1,\ldots\}\to\N$
is def\/ined such that
\[
n!!:=
  \begin{cases}
  n\cdot(n-2)\cdots 2 &\quad \mathrm{if}\ n\ \mathrm{even}\ge 2,
  \\
 n\cdot(n-2)\cdots 1 &\quad \mathrm{if}\ n\ \mathrm{odd}\ge 1,
 \\
  1 &\quad \mathrm{if}\ n=-1,0.
\end{cases}
\]
The \looseness=-1 Jacobi function of the second kind
$Q_\gamma^{(\alpha,\beta)}:\C\setminus(-\infty,1]\to\C$
(cf.~\cite[(10.8.18)]{ErdelyiHTFII}) is def\/ined by
\begin{gather}
 Q_\gamma^{(\alpha,\beta)}(z) :=
\frac{2^{\alpha+\beta+\gamma}\Gamma(\alpha+\gamma+1)\Gamma(\beta+\gamma+1)}
{\Gamma(\alpha+\beta+2\gamma+2)(z-1)^{\alpha+\gamma+1}(z+1)^\beta}
\,{}_2F_1\!\left(
\begin{matrix}
\gamma+1,\alpha+\gamma+1\\
\alpha+\beta+2\gamma+2
\end{matrix}
;\frac{2}{1-z}
\right)\!,\!\!
\label{defnJacobifnsecondkind}
\end{gather}
where $\alpha+\gamma,\beta+\gamma\notin-\N$.
We can derive a relation between the symmetric Jacobi function of the
second kind and the associated Legendre function of the second kind
\begin{gather}
 Q_{n+\nu-1}^{(\mu-\nu+1/2,\mu-\nu+1/2)}(z)
=\frac{2^{\mu-\nu+1/2}\Gamma\left(\mu+n+1/2\right)e^{i\pi(\mu-\nu+1/2)}}
{\Gamma(\nu+n)(z^2-1)^{(\mu-\nu)/2+1/4}}
Q_{n+\mu-1/2}^{\nu-\mu-1/2}(z),
\label{QJACQLEG}
\end{gather}
where $n\in\N_0$,
$\mu\in\C\setminus\big\{{-}\frac12,-\frac32,-\frac52,\ldots\big\}$,
$\nu\in\C\setminus-\N_0$.
The relation (\ref{QJACQLEG}) can be verif\/ied by comparing
(\ref{defnJacobifnsecondkind})
with (\ref{Qdefntwodivide1mz}).

\section{Vilenkin's polyspherical coordinates and the method of trees}
\label{MethodofTrees}

Polyspherical coordinates are hyperspherical coordinates
which are described by a radial coordinate $r\in[0,\infty)$ plus
$(d-1)$-angles which together parametrize points
on $\Si_r^{d-1}$.\footnote{The Riemannian manifold $\Si_r^d$ is def\/ined as the
set of all points in $\R^{d+1}$ such that $x_0^2+\dots+x_d^2=r^2$ $(r>0)$,
with the metric induced from that of the ambient Euclidean space.
We denote the $d$-dimensional hypersphere of unit radius as $\Si^d:=\Si_1^d$.}
We will f\/irst discuss a general procedure for constructing
polyspherical coordinate systems called
the ``method of trees''\footnote{Describing polyspherical coordinate systems in terms of rooted trees
was originally developed in~\cite{VilKuzSmor}
(see also \cite[\S~9.5]{Vilen}, \cite[\S~10.5]{VilenkinKlimyk2}) and has since
been used extensively by others in a variety of contexts
(see for instance \cite{IPSWa,IPSWb,KilMS}).}.
We will give some examples of polyspherical coordinates which will be used
in the rest of the paper and then describe the harmonic separated
product solutions.

\begin{figure}[t]
\centering
\includegraphics[scale=0.95]{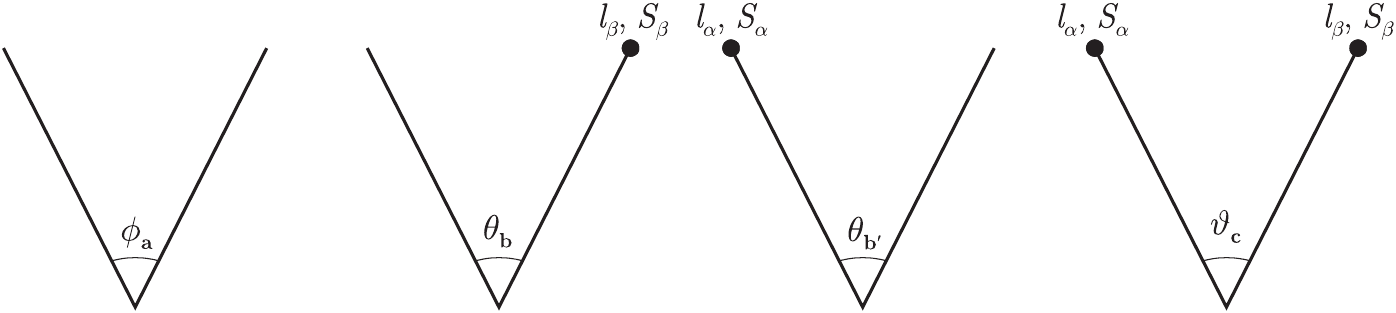}
\caption{This f\/igure shows the possibilities from left to right for branching nodes of
type $\mathbf{a}$, $\mathbf{b}$, ${\mathbf{b}}^\prime$ and~$\mathbf{c}$.
For type $\mathbf{a}$, both branches end on a leaf node.
The angle corresponding to this type of branching node is $\phi_{\mathbf{a}}\in[0,2\pi)$.
For type $\mathbf{b}$, the left
branch ends on a leaf node and the right branch ends on a branching node.
The angle corresponding to this type of branching node is $\theta_{\mathbf{b}}\in[0,\pi]$.
For type ${\mathbf{b}}^\prime$, the left-branch ends on
a branching node and the right branch ends on a leaf node.
The angle corresponding to this type of branching node is
$\theta_{\mathbf{b}^\prime}\in\big[{-}\frac{\pi}{2},\frac{\pi}{2}\big]$.
For type $\mathbf{c}$, both the left and right branches ends
on branching nodes (branching nodes of type $\mathbf{c}$ are only possible for $d\ge 4$).
The angle corresponding to this type of branching node is
$\vartheta_{\mathbf{c}}\in\big[0,\frac{\pi}{2}\big]$.}\label{Fig:top}
\vspace{-2mm}
\end{figure}

In these rooted trees, there are two types of nodes, leaf nodes and
branching nodes.
For a~coordinate system on~$\R^d$, there are $d$-leaf nodes,
each corresponding to the particular
Cartesian component of an arbitrary position vector~$\bfx\in\R^d$.
The branching nodes split into two separate branches,
one up to the left and one up to the right.
Each branch emanating from a branching node will end on either a leaf node
or on another branching node.  There are four possibilities for branching
nodes (see Fig.~\ref{Fig:top}).

Separation of variables
in polyspherical coordinates with $(d-1)$-angles,
for Laplace's equation on $\R^d$ produces $(d-1)$-separation constants, each of
which are called quantum numbers.  The quantum numbers corresponding to these angles
are all integers.
Quantum numbers for a~particular tree label the basis of separable
solutions for Laplace's equation in that particular coordinate system.
With each branching node of the tree, we associate a quantum number as well
as an angle.
The quantum number corresponding to
a $(2\pi)$-periodic (azimuthal) angle is called an azimuthal quantum number.
Each azimuthal angle corresponds to a branching node of type $\mathbf{a}$
and an azimuthal quantum number $m\in\Z$.  A natural consequence of the method of
trees is that there must exist at least one azimuthal angle for each tree, and
therefore also for each polyspherical coordinate system.

Branching nodes of type $\mathbf{b}$, ${\mathbf{b}}^\prime$ and $\mathbf{c}$
(see Fig.~\ref{Fig:top}),
are associated with angles, which in turn are associated with quantum
numbers which we refer to as angular momentum quantum numbers $l\in\N_0$
(see for instance Chapter~10 in~\cite{FanoRau}).
There is always at least one branching node, the root node,
and all branching nodes
correspond to a particular angle and quantum number.  Let us associate each
branching node with an angle and its corresponding quantum number.
These trees parametrize points in polyspherical coordinates as follows.
Starting at the root node, traverse the tree upward until you reach the
leaf node corresponding to~$x_i$.  The parametrization for~$x_i$ is given by
the hyperspherical radius $r$ multiplied by cosine or sine of each angle
encountered as you traverse the tree upward until you reach the leaf node
corresponding to~$x_i$.
If you branch upwards to the left or upwards to the right
at each branching node, multiply by the cosine or sine of the corresponding
angle respectively.  This procedure produces the appropriate transformation
from polyspherical coordinates to Cartesian coordinates.

There are large numbers of equivalent
trees and an even larger
number of total trees, each with their own specif\/ic
polyspherical coordinate system.
The enumeration of these trees are characterized as follows.
For $d,b_d\in\N$, let $b_d$ be the total number of trees.
Then $b_1=1$ is the number of possible
1-branch trees.  In our context, a 1-branch tree does not exist in isolation.
The following recurrence relation gives the total number of trees
for arbitrary dimension
\begin{gather}
  b_d=\sum_{i=1}^{d-1}b_i b_{d-i}.
\label{recurb}
\end{gather}
 Using the recurrence relation (\ref{recurb}), the
f\/irst few elements of the sequence are given by
\[
\left(b_d:d\in\{2,\ldots,13\}\right)
=(1,\,2,\,5,\,14,\,42,\,132,\,429,\,1430,\,4862,\,16796,\,58786,\,208012).
\]
The total number of trees are given in terms of
the Catalan numbers $C_n$ (see for
instance Sloane integer sequence
A000108 \cite{Sloane} or p.~200 in~\cite{Stanley}),
i.e.,\
$b_d=C_{d-1}=\frac{1}{d}{2d-2 \choose  d-1}$,
where
${n \choose k}=\frac{n!}{k!(n-k)!}$
is the binomial coef\/f\/icient for $k,n\in\N_0$ with $0\le k\le n$.
If $a_d\in\N$ is the total number of equivalence classes for equivalent trees
for a given dimension
(determined by a left-right symmetry in the topology of the trees),
then $a_d$ satisf\/ies the following
recurrence relation
\begin{gather}
a_d=
\begin{cases}
\displaystyle \sum_{i=1}^{\lfloor d/2\rfloor}
a_ia_{d-i}, &\quad\mathrm{if}\ d \ \mathrm{odd,}
\vspace{1mm}\\
\displaystyle \sum_{i=1}^{d/2-1}a_i a_{d-i}+\frac12a_{d/2}\left(a_{d/2}+1\right),
&\quad\mathrm{if}\ d \ \mathrm{even.}
\end{cases}
\label{recura}
\end{gather}
Using the recurrence relation (\ref{recura}),
the f\/irst few elements of the sequence are given by
\[
\left(a_d:d\in\{2,\ldots,13\}\right)
=(1,\,1,\,2,\,3,\,6,\,11,\,23,\,46,\,98,\,207,\,451,\,983).
\]
These are given in terms of the Wedderburn--Etherington numbers
(see for instance
Sloane integer sequence A001190~\cite{Sloane}).
We use a left-to-right recursive naming language for our trees
based on a depth-f\/irst search (see~\cite[pp.~540--549]{Cormenetal}).  This naming language is given by listing the types
of branching nodes available in a particular tree.

In a polyspherical coordinate system the Euclidean distance between two
points~(\ref{algebraicdist})
can also be given as
\begin{gather}
\|\bfx-\bfxp\|=\sqrt{2rr'}\left[\frac{r^2+r'^2}{2rr'}-\cos\gamma\right]^{1/2},
\label{distsph}
\end{gather}
where the separation angle $\gamma\in[0,\pi]$
is def\/ined through
the relation
\begin{gather}
\cos\gamma:=\frac{(\bfx,\bfxp)}{\|\bfx\|\|\bfxp\|},
\label{separationangle}
\end{gather}
using the Euclidean inner product and norm
(cf.~(\ref{eucinnerprod})).
The method of trees constructs the cosine of the
separation angle~(\ref{separationangle})
in a direct manner.
The cosine of the separation angle will be given by the sum of $d$-terms,
each corresponding to a leaf node of the tree.  There is a unique
path starting from the root node to each
leaf node. It is
\[
\cos\gamma=\sum_{i=1}^d \prod_{j=1}^{N_i} A_{i,j}(\psi_{i,j})A_{i,j}(\psi_{i,j}^\prime),
\]
where $N_i$ is the number of branching nodes encountered from the root node to the
leaf node, $A_{i,j}:\R\to[-1,1]$ is either the trigonometric cosine or sine function
depending respectively on whether the left branch or right branch
is chosen respectively,
and $\psi_{i,j}\in\R$ is
the angle corresponding to the $j$th branching node for each $i$th leaf
node.  The formula for the cosine of the separation angle is unique for each tree.

\subsection{Examples of Vilenkin's polyspherical coordinate systems}

\begin{figure}[t]
\centering
\includegraphics[scale=0.95]{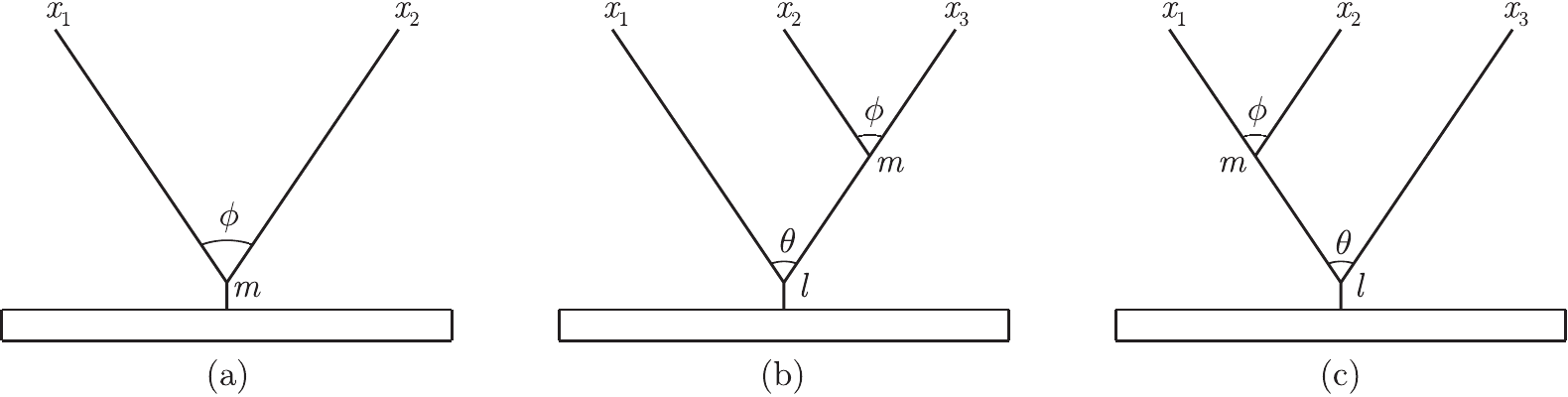}
\caption{Tree diagrams for two and three dimensional polyspherical coordinate systems of type:
(a)~$\mathbf{a}$,
(b)~$\mathbf{ba}$,
(c)~${\mathbf{b}}^\prime \mathbf{a}$.}
\label{Fig:tree23}\vspace{-2mm}
\end{figure}

\begin{figure}[t]
\centering
\includegraphics[scale=0.93]{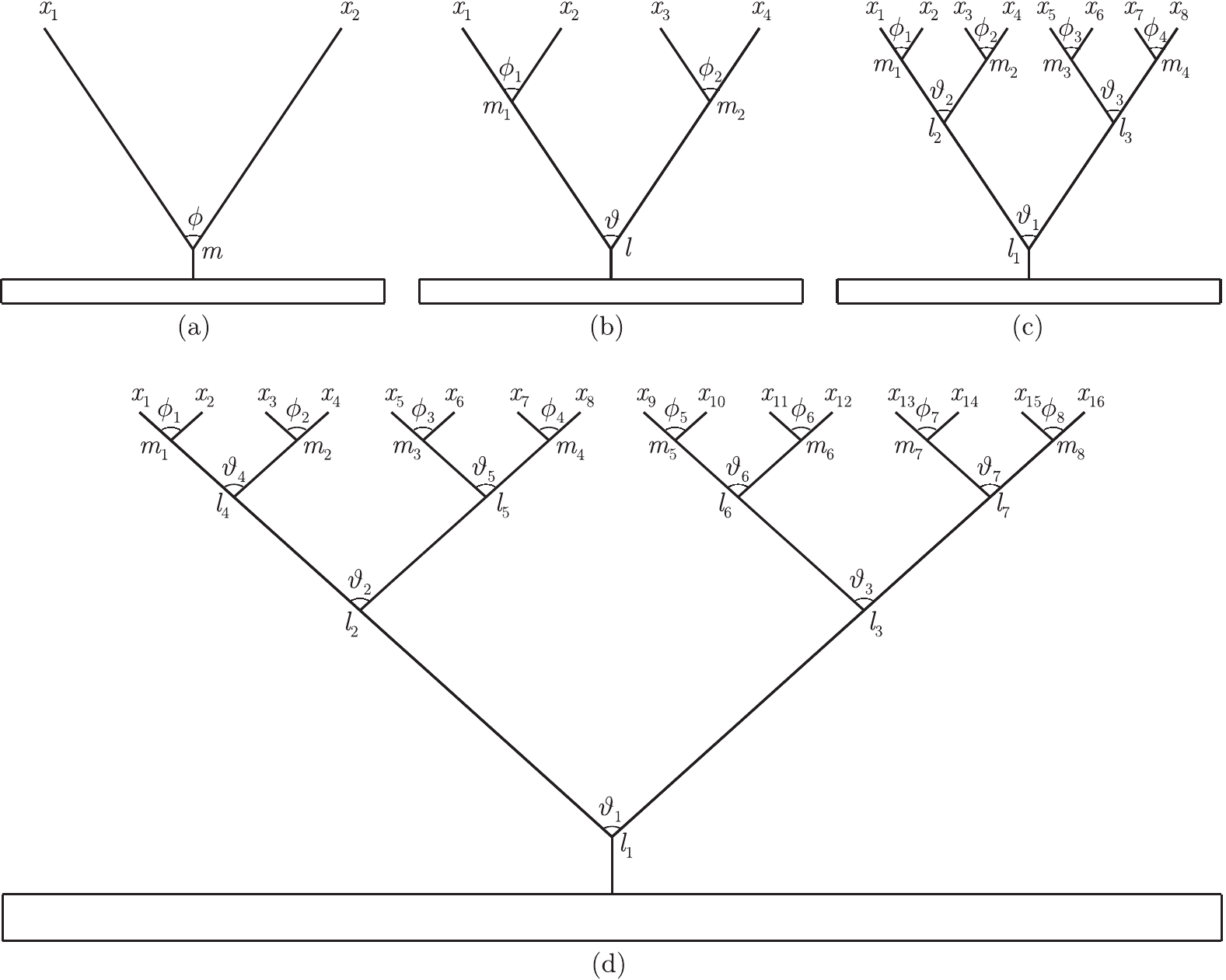}
\caption{This f\/igure is a tree diagram for polyspherical generalized Hopf
coordinates of type ${\mathbf{V}}_{2^q}$ on~$\R^{2^{q}}$ with $q=1,2,3,4$
for (a), (b), (c), (d) respectively.
The f\/irst $(2^{q-1}-1)$-branching nodes are of type $\mathbf{c}$ which correspond to the angles
$\vartheta_i\in\big[0,\frac{\pi}{2}\big]$
and quantum numbers~$l_i\in\N_0$.  The following $(2^q)$-branching nodes are of type
${\mathbf{a}}$ which correspond to the angles $\phi_i\in[0,2\pi)$ and quantum numbers $m_i\in\Z$.
These coordinates correspond to transformation~(\ref{genHopf}).}
\label{Fig:genHopf}
\end{figure}

The simplest example of a polyspherical
coordinate system on $\R^d$
occurs for $d=2$ (polar coordinates) where there is one branching
node (the root node) and two leaf nodes (see
Figs.~\ref{Fig:tree23}a and~\ref{Fig:genHopf}a).
The left-branch ends on the
leaf node corresponding
to $x_1$ and the right branch ends on the leaf node corresponding to $x_2$,
i.e.,
\begin{gather}
x_1 = r\cos\phi, \qquad
x_2 = r\sin\phi ,
\label{hypsph2}
\end{gather}
where $r\in[0,\infty)$ and $\phi\in[0,2\pi)$.
We refer to this tree as type $\mathbf{a}$.
The cosine of the separation angle is given by{\samepage
\begin{gather}
\cos\gamma=\cos(\phi-\phi^\prime),
\label{cosgamma2d}
\end{gather}
and corresponding to the angle $\phi$ is the azimuthal quantum number $m\in\Z$
(see Fig.~\ref{Fig:tree23}a).}

In $d=3$ there are two possible topological trees, each
corresponding to one of two dif\/ferent trees.
The f\/irst tree
(see Fig.~\ref{Fig:tree23}b)
corresponds to
the coordinate system
\begin{gather}
x_1 = r\cos\theta,\qquad
x_2 = r\sin\theta\cos\phi, \qquad
x_3 = r\sin\theta\sin\phi,
\label{hypsph3a}
\end{gather}
where $\theta\in[0,\pi]$, $\phi\in[0,2\pi)$,
i.e.,
{\em standard spherical coordinates}.
This is a tree of type $\mathbf{ba}$.
The cosine of the separation angle is given by
\begin{gather*}
\cos\gamma=\cos\theta\cos\theta^\prime+\sin\theta\sin\theta^\prime\cos(\phi-\phi^\prime),
\end{gather*}
and corresponding to the angles $\theta\in[0,\pi]$ and $\phi\in[0,2\pi)$ are
the quantum numbers
$l\in\N_0$ and $m\in\Z$ respectively.
The second tree
(see Fig.~\ref{Fig:tree23}c)
is of type ${\mathbf{b}}^\prime \mathbf{a}$ and is equivalent to the f\/irst.

\begin{figure}[t]
\centering
\includegraphics[scale=0.95]{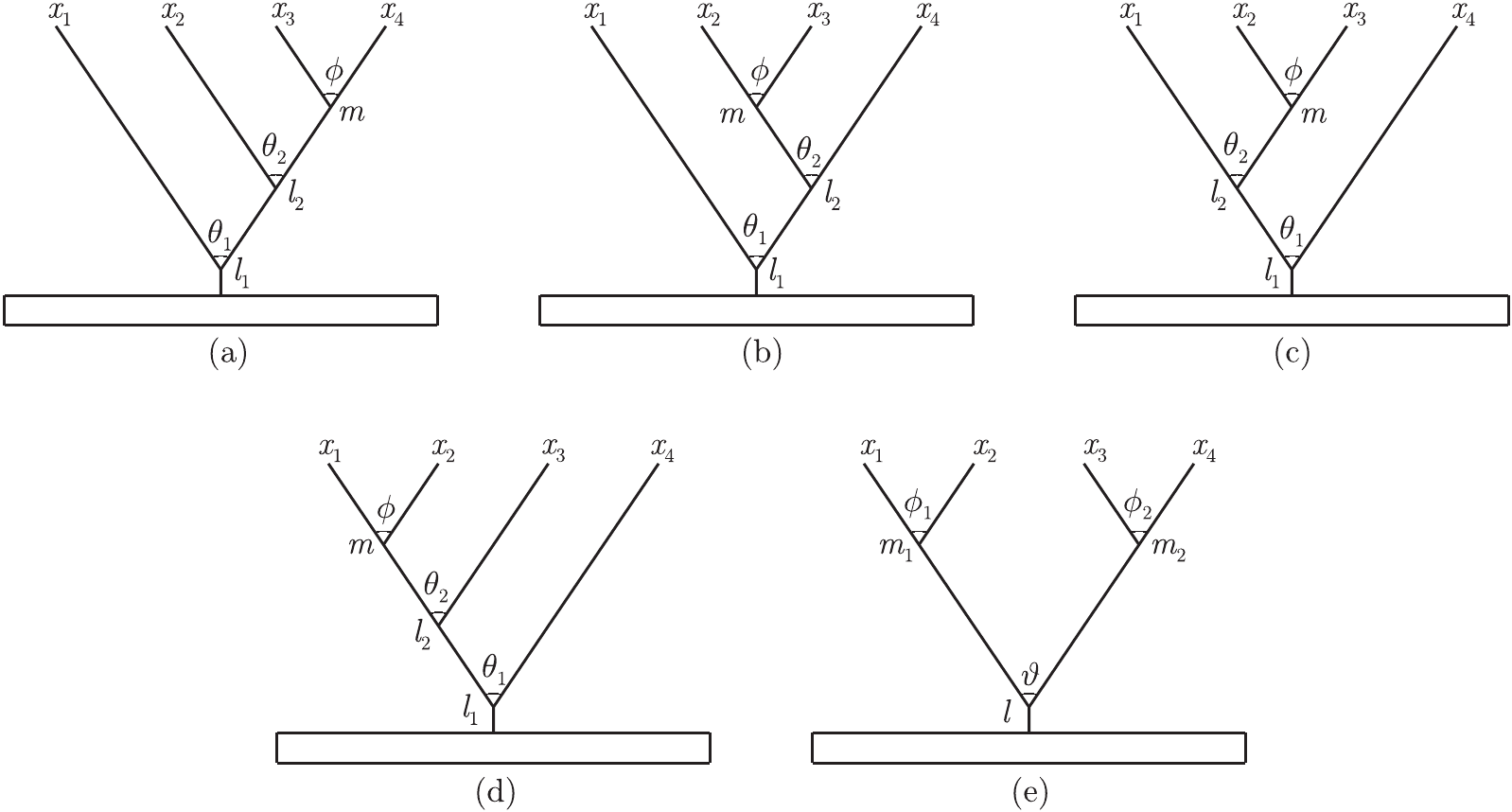}
\caption{Tree diagrams for four dimensional polyspherical coordinate systems of type:
(a)~${\mathbf{b}}^2 \mathbf{a}$,
(b)~${\mathbf{bb}}^\prime \mathbf{a}$,
(c)~${\mathbf{b}}^\prime \mathbf{ba}$,
(d)~${\mathbf{b}}^{\prime 2} \mathbf{a}$,
(e)~${\mathbf{ca}}^2$.}
\label{Fig:tree4}
\end{figure}

\begin{figure}[t]
\centering
\includegraphics[scale=0.95]{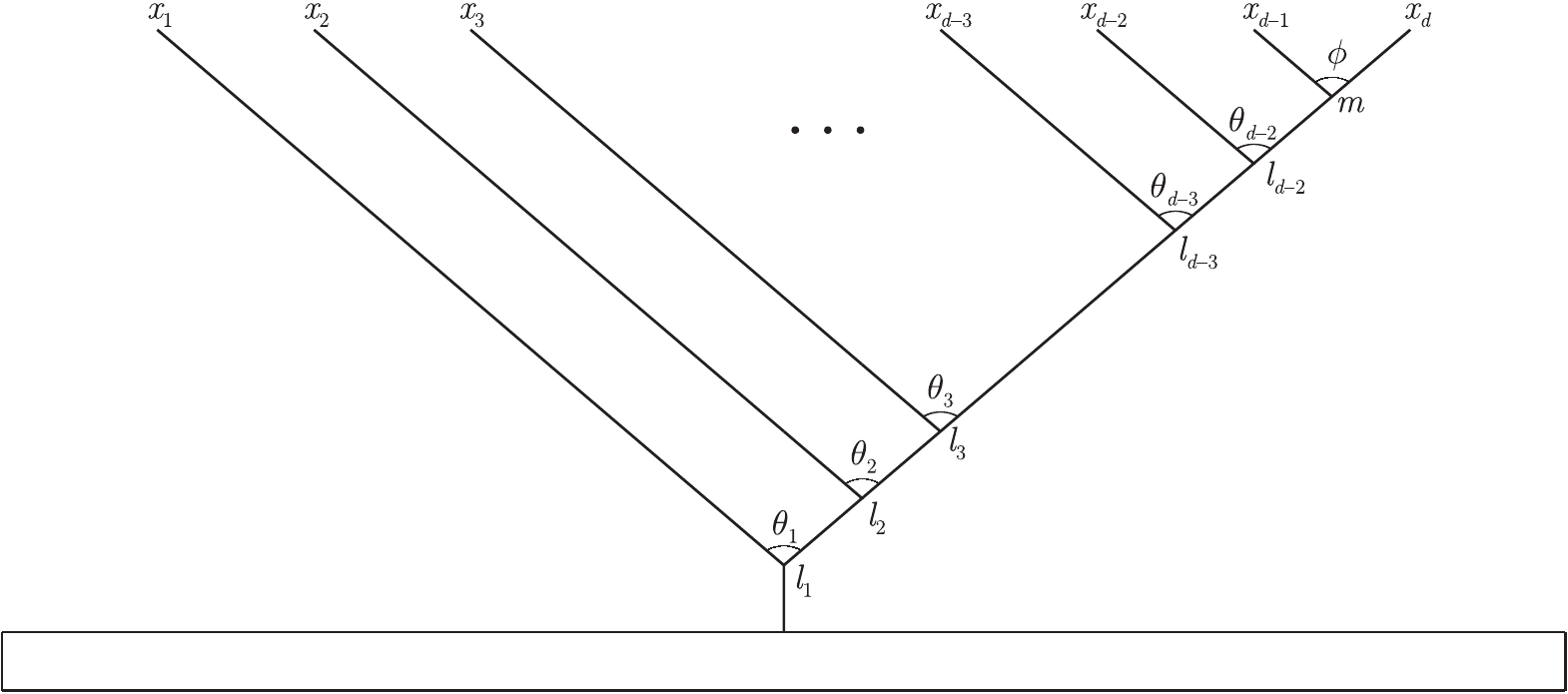}
\caption{Tree diagram for $d$-dimensional standard polyspherical coordinates of type
${\mathbf{b}}^{d-2} {\mathbf{a}}$ (standard polyspherical coordinates).}
\label{Fig:stansph}
\end{figure}

In $d=4$ there are f\/ive possible topological trees
for polyspherical coordinates.
The f\/irst tree
(see Figs.~\ref{Fig:tree4}a and~\ref{Fig:stansph})
corresponds to the coordinate system
\begin{alignat}{3}
& x_1 = r\cos\theta_1, \qquad && x_2 = r\sin\theta_1\cos\theta_2, & \nonumber\\
& x_3 = r\sin\theta_1\sin\theta_2\cos\phi, \qquad && x_4 = r\sin\theta_1\sin\theta_2\sin\phi, & \label{typeb2a}
\end{alignat}
where $\theta_1,\theta_2\in[0,\pi]$.
This tree is of type ${\mathbf{b}}^2{\mathbf{a}}$, and
the cosine of the separation angle is given by{\samepage
\begin{gather*}
\cos\gamma=
\cos\theta_1\cos\theta_1^\prime
+\sin\theta_1\sin\theta_1^\prime
\left(
\cos\theta_2\cos\theta_2^\prime+\sin\theta_2
\sin\theta_2^\prime\cos(\phi-\phi^\prime)
\right).
\end{gather*}
The second, third and fourth trees
(see Figs.~\ref{Fig:tree4}b,~\ref{Fig:tree4}c,~\ref{Fig:tree4}d) are equivalent to the f\/irst.}

The f\/ifth tree
(see Figs.~\ref{Fig:tree4}e and~\ref{Fig:genHopf}b)
corresponds
to {\em Hopf coordinates},
\begin{alignat}{3}
& x_1 = r\cos\vartheta\cos\phi_1,\qquad &&
x_2 = r\cos\vartheta\sin\phi_1, &\nonumber\\
& x_3 = r\sin\vartheta\cos\phi_2,\qquad &&
x_4 = r\sin\vartheta\sin\phi_2, & \label{typeca2}
\end{alignat}
where $\vartheta\in\big[0,\frac{\pi}{2}\big]$ and $\phi_1,\phi_2\in[0,2\pi)$.
This tree is of type ${\mathbf{ca}}^2$.
The cosine of the separation angle is given by
\begin{gather}
\cos\gamma=
\cos\vartheta\cos\vartheta^\prime\cos(\phi_1-\phi_1^\prime)
+\sin\vartheta\sin\vartheta^\prime\cos(\phi_2-\phi_2^\prime).
\label{cosgammaca2}
\end{gather}

There are many choices for polyspherical coordinates on $\R^d$,
suitably def\/ined for any number of dimensions $d\ge 2$.
The simplest example of a polyspherical coordinate system on $\R^d$
(which generalizes type ${\mathbf{ba}}$ and ${\mathbf{b}}^2{\mathbf{a}}$)
are what we refer to as {\it standard polyspherical coordinates}
(see Fig.~\ref{Fig:stansph}).
These are
\begin{gather}
x_{1} = r\cos\theta_1,\nonumber\\
x_{2} = r\sin\theta_1\cos\theta_2,\nonumber\\
x_{3} = r\sin\theta_1\sin\theta_2\cos\theta_3,\nonumber\\
\cdots \cdots\cdots\cdots\cdots\cdots\cdots\cdots\nonumber\\
x_{d-2} = r\sin\theta_1\cdots\sin\theta_{d-3}\cos\theta_{d-2},\nonumber\\
x_{d-1} = r\sin\theta_1\cdots\sin\theta_{d-3}\sin\theta_{d-2}\cos\phi,\nonumber\\
x_{d} = r\sin\theta_1\cdots\sin\theta_{d-3}\sin\theta_{d-2}\sin\phi,\label{standardsph}
\end{gather}
where $\theta_i\in[0,\pi]$ for
$1\le i \le d-2$ and
$\phi\in[0,2\pi)$.
Using our naming procedure, this
tree is of type ${\mathbf{b}}^{d-2}{\mathbf{a}}$.
In these coordinates
the cosine of the separation angle is given by
\[
 \cos \gamma=
\sum_{i=1}^{d-2}\cos\theta_i{\cos\theta_i'}
\prod_{j=1}^{i-1}\sin\theta_j{\sin\theta_j'}
+\cos(\phi-\phi^\prime)\prod_{i=1}^{d-2}\sin\theta_i{\sin\theta_i'}.
\]

Another example of a polyspherical coordinate system which is valid for
large dimensions is what we will refer to as
{\it generalized Hopf coordinates}
(see Fig.~\ref{Fig:genHopf}).
These coordinates, valid on~$\R^{2^q}$ for~$q\ge 1$,
generalize two-dimensional polar coordinates (\ref{hypsph2}, type ${\mathbf{a}}$) and
four-dimensional Hopf coordinates~(\ref{typeca2}, type ${\mathbf{ca}}^2$).
These coordinates are unique in that they correspond
to the only trees which contain only themselves
in their equivalence class (see~(\ref{recura})).
These coordinate systems have separated harmonic eigenfunctions which
are given in terms of complex
exponentials, and for $q\ge 2$, non-symmetric Jacobi polynomials
(see Fig.~\ref{Fig:genHopf}).

This coordinate system is suitably def\/ined for dimensions $d=2^q$ for $q\ge 1$.
The
transformation formulae to Cartesian coordinates
are given by
\begin{gather}
x_{1} = r\cos\vartheta_1\cos\vartheta_2\cos\vartheta_4\cos\vartheta_8\cdots\cos\vartheta_{2^{q-2}}\cos\phi_1,\nonumber\\
x_{2} = r\cos\vartheta_1\cos\vartheta_2\cos\vartheta_4\cos\vartheta_8\cdots\cos\vartheta_{2^{q-2}}\sin\phi_1,\nonumber\\
\cdots\cdots\cdots\cdots\cdots\cdots\cdots\cdots\cdots\cdots\cdots\cdots\cdots\cdots\cdots\cdots  \nonumber\\
x_{2^{q-1}-1} = r\cos\vartheta_1\sin\vartheta_2\sin\vartheta_5\sin\vartheta_{11}\cdots
\sin\vartheta_{3\cdot 2^{q-3}-1}\cos\phi_{2^{q-2}},\nonumber\\
x_{2^{q-1}} = r\cos\vartheta_1\sin\vartheta_2\sin\vartheta_5\sin\vartheta_{11}\cdots
\sin\vartheta_{3\cdot 2^{q-3}-1}\sin\phi_{2^{q-2}},\nonumber\\
x_{2^{q-1}+1} = r\sin\vartheta_1\cos\vartheta_3\cos\vartheta_6\cos\vartheta_{12}\cdots
\cos\vartheta_{3\cdot 2^{q-3}}\cos\phi_{2^{q-2}+1},\nonumber\\
x_{2^{q-1}+2} = r\sin\vartheta_1\cos\vartheta_3\cos\vartheta_6\cos\vartheta_{12}\cdots
\cos\vartheta_{3\cdot 2^{q-3}}\sin\phi_{2^{q-2}+1},\nonumber\\
\cdots\cdots\cdots\cdots\cdots\cdots\cdots\cdots\cdots\cdots\cdots\cdots\cdots\cdots\cdots\cdots  \nonumber\\
x_{2^{q}-1} = r\sin\vartheta_1\sin\vartheta_3\sin\vartheta_7\sin\vartheta_{17}\cdots
\sin\vartheta_{2^{q-1}-1}\cos\phi_{2^{q-1}},\nonumber\\
x_{2^{q}} = r\sin\vartheta_1\sin\vartheta_3\sin\vartheta_7\sin\vartheta_{17}\cdots
\sin\vartheta_{2^{q-1}-1}\sin\phi_{2^{q-1}},
\label{genHopf}
\end{gather}
where $\vartheta_i\in\big[0,\frac{\pi}{2}\big]$ for
$1\le i \le 2^{q-1}-1$
and $\phi_i\in[0,2\pi)$ for
$1\le i\le 2^{q-1}$.
Using our naming procedure, these coordinates are of type{\samepage
\[
{\mathbf{V}}_{2^q}={\mathbf{c}}{\mathbf{V}}_{2^{q-1}}{\mathbf{V}}_{2^{q-1}},
\]
where $q\in\N$, with type ${\mathbf{V}}_2= \mathbf{a}$ (polar) coordinates on $\R^2$
(see Figs.~\ref{Fig:tree23}a and~\ref{Fig:genHopf}a).}

The cosine of the separation angle in this coordinate system may be given as follows.
Def\/ine the symbol ${}_q{\sf G}_s^r\in[-1,1]$, where $0\le s\le q$
and $1\le r \le 2^{q}-1$,
by the recursive formula
\[
 {}_{q}{\sf G}_s^r=
\cos\vartheta_{r-1+2^{q-s}}
\cos\vartheta_{r-1+2^{q-s}}'
\ {}_{q}{\sf G}_{s-1}^{2r-1}
+
\sin\vartheta_{r-1+2^{q-s}}
\sin\vartheta_{r-1+2^{q-s}}'
\ {}_{q}{\sf G}_{s-1}^{2r},
\]
with ${}_{q}{\sf G}_0^i=1$. Then the cosine of the separation angle is
given by
\[
\cos\gamma={}_{q}{\sf G}_q^1
\]
(cf.~(\ref{cosgamma2d}), (\ref{cosgammaca2})).
Note that through the identif\/ication
$\phi_i=\vartheta_{i-1+2^{q-1}}$, where
$1\le i\le 2^{q-1}$,
then
${}_{q}{\sf G}_1^i=\cos(\phi_i-\phi_i')$. Thus, this shows one may stop this recursion at $s=1$.

\subsection{Hyperspherical harmonics in polyspherical coordinates}
\label{Hypersphericalharmonics}

The eigenfunction expansions for a power-law fundamental solution of
the polyharmonic equation in polyspherical coordinates
can be derived using a Gegenbauer polynomial expansion for the
relevant kernel (see Corollary~\ref{MYCOROLLARY}) in conjunction
with the addition theorem for hyperspherical harmonics.  This addition
theorem is given by
\begin{gather}
C_n^{d/2-1}(\cos\gamma)
=\frac
{2
(d-2)
\pi^{d/2}
}
{
(2n+d-2)
\Gamma(d/2)
}
\sum_{K}
Y_n^K
(\wbfx)
\overline{Y_n^K
(\wbfxp)
}
\label{addthmhypsph}
\end{gather}
(for a proof see 
\cite{WenAvery};
see also \S~10.2.1 in
\cite{FanoRau}),
where $K$ stands for a set of $(d-2)$-quantum numbers
identifying harmonics
for a given value of ${n}\in\N_0$, and $\cos\gamma$ is the cosine of the
separation
angle between two arbitrary
vectors $\bfx,\bfxp\in\R^d$ (see (\ref{separationangle})).
The functions
$Y_{{n}}^{K}:\Si^{d-1}\to\C$ are the normalized hyperspherical harmonics.
Normalization of the hyperspherical harmonics is achieved through
\[
\int_{\Si^{d-1}}Y_n^K(\wbfx)\overline{Y_n^K(\wbfx)}d\Omega=1,
\]
where $d\Omega$ is the Riemannian volume measure on $\Si^{d-1}$.

The general basis functions that one obtains by putting coordinates on the $d$-dimensional
unit hypersphere $\Si^{d-1}$, can be specif\/ied as solutions to
the angular part of Laplace's
equation
on~$\R^{d}$.  These correspond
to separated solutions of Laplace's equation,
using the Laplace--Beltrami ope\-rator
on the hypersphere $\Si^{d-1}$.
The following numbers are associated with each
branching node $m\in\Z$, $l,l_\alpha,l_\beta\in\N_0$ (see Fig.~\ref{Fig:top}).
The number of vertices above each branching node
$l_\alpha$ and $l_\beta$ are represented by $S_\alpha$ and $S_\beta$ respectively.

The following separated factors of eigenfunctions are generated at each
branching node for
normalized hyperspherical harmonics in polyspherical coordinates
using the method of trees
(see for instance \cite[(2.3)--(2.6)]{IPSWb},
\cite[\S~10.5.3]{VilenkinKlimyk2}):
\begin{itemize}\itemsep=0pt

\item Type ${\mathbf{a}}$:
\begin{gather*}
\Psi_m(\phi_{\mathbf{a}})=\frac{1}{\sqrt{2\pi}}e^{im\phi_{\mathbf{a}}},\qquad m\in\Z, \qquad\phi_{\mathbf{a}}\in[0,2\pi).
\end{gather*}

\item Type ${\mathbf{b}}$:
\begin{gather}
\Psi_{n,l_\beta}^\alpha(\theta_{\mathbf{b}})=N_n^{\alpha,\alpha}(\sin\theta_{\mathbf{b}})^{l_\beta}
P_n^{(\alpha,\alpha)}(\cos\theta_{\mathbf{b}}),\nonumber\\
 n=l-l_\beta,\qquad   \alpha=l_\beta+\frac{S_\beta}{2}, \qquad n\in\N_0,\qquad \theta_{\mathbf{b}}\in[0,\pi].
\label{cellb}
\end{gather}

\item Type ${\mathbf{b}}^\prime $:
\begin{gather}
\Psi_{n,l_\alpha}^\beta(\theta_{{\mathbf{b}}^\prime})=N_n^{\beta,\beta}
(\cos\theta_{{\mathbf{b}}^\prime})^{l_\alpha}
P_n^{(\beta,\beta)}(\sin\theta_{{\mathbf{b}}^\prime}),\nonumber\\
 n=l-l_\alpha,\qquad
 \beta=l_\alpha+\frac{S_\alpha}{2}, \qquad n\in\N_0,
\qquad\theta_{{\mathbf{b}}^\prime} \in\left[-\frac{\pi}{2},\frac{\pi}{2}\right]
.
\label{cellbp}
\end{gather}

\item Type ${\mathbf{c}}$:
\begin{gather}
\Psi_{n,l_\alpha,l_\beta}^{\alpha,\beta}
(\vartheta_{\mathbf{c}})=
2^{(\alpha+\beta)/2+1}N_n^{\alpha,\beta}
(\sin\vartheta_{\mathbf{c}})^{l_\beta}
(\cos\vartheta_{\mathbf{c}})^{l_\alpha}
P_n^{(\beta,\alpha)}(\cos 2\vartheta_{\mathbf{c}}),\nonumber\\
n=\frac12\left(l-l_\alpha-l_\beta\right), \qquad   \alpha=l_\alpha+\frac{S_\alpha}{2}, \qquad
 \beta=l_\beta+\frac{S_\beta}{2}, \nonumber\\ n\in\N_0,
\qquad\vartheta_{\mathbf{c}} \in\left[0,\frac{\pi}{2}\right].
\label{cellc}
\end{gather}
\end{itemize}

We refer to the quantum number $n\in\N_0$
in
(\ref{cellb})--(\ref{cellc}) as the {\it surrogate}
quantum number to $l\in\N_0$.

Notice that the eigenfunctions for branching nodes of
type $\mathbf{b}$ and ${\mathbf{b}}^\prime$ can be expressed in
terms of Gegenbauer polynomials using~(\ref{gegenjac}).
Therefore we can re-write (\ref{cellb}) as
\begin{gather}
\Psi_{n,l_\beta}^\alpha(\theta_{\mathbf{b}})=
\frac{(2\alpha)!}{\Gamma\left(\alpha+1\right)}\sqrt{\frac{(2\alpha+2n+1)n!}{2^{2\alpha+1}(2\alpha+n)!}}
(\sin\theta_{\mathbf{b}})^{l_\beta}
C_n^{\alpha+1/2}(\cos\theta_{\mathbf{b}}),
\label{psithetab}
\end{gather}
and (\ref{cellbp}) as{\samepage
\begin{gather*}
\Psi_{n,l_\alpha}^\beta(\theta_{{\mathbf{b}}^\prime})=
\frac{(2\beta)!}{\Gamma\left(\beta+1\right)}\sqrt{\frac{(2\beta+2n+1)n!}{2^{2\beta+1}(2\beta+n)!}}
(\sin\theta_{{\mathbf{b}}^\prime})^{l_\alpha}
C_n^{\beta+1/2}(\cos\theta_{{\mathbf{b}}^\prime}).
\end{gather*}
Note that even though $\alpha$, $\beta$ are not necessarily integers,
$2\alpha, 2\beta\in\N_0$.}

With the simplest example, polar coordinates (see Fig.~\ref{Fig:tree23}a,
transformation (\ref{hypsph2})),
the normalized harmonics are given by
\[
Y_m(\phi)=\frac{e^{im\phi}}
{\sqrt{2\pi}}.
\]

In $d=3$ there are two equivalent polyspherical coordinate
systems, that of type
$\mathbf{ba}$ (\ref{hypsph3a}) and
${\mathbf{b}}^\prime \mathbf{a}$.
Since they are equivalent coordinate systems, we treat only the f\/irst.
This tree has two branching nodes.
Using~(\ref{cellb})
we see that $\alpha=m$, $n=l-m$, $l_\beta=m$, and
$S_\beta=0$ since there are no vertices above the branching
node~$m$.
Through reduction and multiplication by the $\phi$
eigenfunction, the normalized
spherical harmonics are
\begin{gather}
Y_{l,m}(\theta,\phi)=(-1)^m
\sqrt{
\frac{2l+1}{4\pi}\frac{(l-m)!}{(l+m)!}} {\rm P}_l^m(\cos\theta)e^{im\phi}.
\label{typeba}
\end{gather}
Notice we have used (\ref{GegenLeg})
with (\ref{psithetab}) to reduce the Gegenbauer polynomial
to a Ferrers function of the f\/irst kind.
The functions $Y_{l,m}:[0,\pi]\times[0,2\pi)\to\C$ are
called {\em standard spherical harmonics.}

In $d=4$ we consider type~${\mathbf{b}}^2{\mathbf{a}}$ (see Fig.~\ref{Fig:tree4}a) coordinates,
whose normalized hyperspherical harmonics are
\begin{gather*}
 Y_{l_1,l_2,m}(\theta_1,\theta_2,\phi)=\frac{(-1)^m(2l_2)!!}{\pi}
\sqrt{\frac{(2l_2+1)(l_1+1)(l_1-l_2)!(l_2-m)!}{2(l_1+l_2+1)!(l_2+m)!}}\nonumber\\
\hphantom{Y_{l_1,l_2,m}(\theta_1,\theta_2,\phi)=}{}
{}\times(\sin\theta_1)^{l_2}
C_{l_1-l_2}^{l_2+1}(\cos\theta_1)
{\rm P}_{l_2}^m(\cos\theta_2)e^{im\phi}.
\end{gather*}
In type ${\mathbf{ca}}^2$
(see Fig.~\ref{Fig:tree4}e), the normalized hyperspherical harmonics are
\begin{gather*}
  Y_{l,m_1,m_2}(\vartheta,\phi_1,\phi_2)=\frac{
e^{i(m_1\phi_1+m_2\phi_2)}
}{\pi}
{  \sqrt{
\frac{l+1}{2}\frac{\left[\frac12(l+|m_1|+|m_2|\right]!\left[\frac12(l-|m_1|-|m_2|)\right]!}
{\left[\frac12(l-|m_1|+|m_2|)\right]!\left[\frac12(l+|m_1|-|m_2|)\right]!}}}\nonumber\\
\hphantom{Y_{l,m_1,m_2}(\vartheta,\phi_1,\phi_2)=} {}\times
(\sin\vartheta)^{|m_2|}
(\cos\vartheta)^{|m_1|}
P_{(l-|m_1|-|m_2|)/2}^{(|m_2|,|m_1|)}(\cos2\vartheta),
\end{gather*}
with the restriction to the parameter space given by
$\frac{1}{2}(l-|m_1|-|m_2|)\in\N_0$.  Using the surrogate quantum number $n\in\N_0$,
def\/ined in terms of $l\in\N_0$ such that $2n=l-|m_1|-|m_2|$, we can more conveniently
express the normalized hyperspherical harmonics in type ${\mathbf{ca}}^2$ coordinates,
namely
\begin{gather*}
  Y_{n,m_1,m_2}(\vartheta,\phi_1,\phi_2)=\frac{e^{i(m_1\phi_1+m_2\phi_2)}}{\pi}
\sqrt{\frac
{(2n+|m_1|+|m_2|+1)(n+|m_1|+|m_2|)!n!}
{2(n+|m_1|)!(n+|m_2|)!}} \\
\hphantom{Y_{n,m_1,m_2}(\vartheta,\phi_1,\phi_2)=}{}
\times(\sin\vartheta)^{|m_2|}
(\cos\vartheta)^{|m_1|}
P_n^{(|m_2|,|m_1|)}(\cos 2\vartheta).
\end{gather*}

For arbitrary dimensions, we can use standard polyspherical coordinates~(\ref{standardsph}) to construct the normalized hyperspherical harmonics.
The polyspherical harmonics corresponding to this coordinate system are basis
functions for the irreducible representations of~$O(d)$
(see~\cite{Vilen}).  In terms of these
coordinates, the normalized hyperspherical harmonics are
\begin{gather}
Y_{l}^K
(\widehat{\mathbf{x}})
=\frac{e^{im\phi}}{\sqrt{2\pi}}
\prod_{j=1}^{d-2}\Theta_j^d(l_{j},l_{j+1};\theta_j),
\label{standardhypersphericalharmonic}
\end{gather}
where $ \widehat{\mathbf{x}}\in\Si^{d-1}$,
$K=\{l_2,l_3,\ldots,l_{d-1}\}$,
$l=l_1\ge l_2\ge l_3\ge\!\cdots\!\ge l_{d-3}\ge l_{d-2}=\ell
\ge l_{d-1} = |m| \ge 0$,
and $\Theta_j^d:\N_0^2\times[0,\pi]\to\R$ is def\/ined by
\begin{gather}
 \Theta_j^d(l_{j},l_{j+1};\theta_j) :=
{  \frac{\Gamma\left(l_{j+1}+{\frac{d-j+1}{2}}\right)}
{2l_{j+1}+d-j-1}}
\sqrt{
\frac{2^{2l_{j+1}+d-j-1}
(2l_j+d-j-1)
(l_{j}-l_{j+1})!
}{\pi(l_j+l_{j+1}+d-j-2)!}}\nonumber\\
\hphantom{\Theta_j^d(l_{j},l_{j+1};\theta_j) :=}{}
\times
(\sin\theta_j)^{l_{j+1}}
C_{l_{j}-l_{j+1}}^{l_{j+1}+(d-j-1)/2}(\cos\theta_j).
\label{stansphharmoncis}
\end{gather}
The computation of
(\ref{stansphharmoncis})
is a straightforward consequence of~(\ref{psithetab}),
and doing
the proper node counting for $S_\beta$, in the tree depicted in
Fig.~\ref{Fig:stansph}.
The normalized spherical harmonics~(\ref{typeba}) are the simplest example of these harmonics.

In generalized Hopf coordinates of type ${\mathbf{V}}_{2^p}$ coordinates with $p\ge 1$
(see Fig.~\ref{Fig:genHopf}),
the normalized hyperspherical harmonics
$Y_{l_{1}}^K:\Si^{2^{p}-1}\to\C$
can be given more conveniently using the surrogate quantum
numbers $n_q\in\N_0$ which are connected to
$l_q\in\N_0$ through the relation $l_q=l_\alpha+l_\beta+2n_q$ such that $1\le q \le 2^{p-1}-1$.
The orthonormal polyspherical harmonics, which can be generated using the method of trees, are
\begin{gather}
Y_{l_{1}}^K
(\widehat{\mathbf{x}})=
\frac{e^{i\left(m_{1}\phi_{1}+\cdots+m_{2^{p-1}}\phi_{2^{p-1}}\right)}}{\sqrt{2}\pi^{2^{p-2}}}
\prod_{j=1}^{p-1} \prod_{s=1}^{2^{j-1}} \Upsilon_q^p
\left(
\begin{matrix}
n_{q}\\
l_{\alpha},l_{\beta}
\end{matrix};\vartheta_{q} \right),
\label{genHopfcoordinatesthis}
\end{gather}
where
$K=\{l_2,\ldots,l_{2^{p-1}-1},m_1,\ldots,m_{2^{p-1}}\}$ and
$\Upsilon_q^p:\N_0^3\times[0,\frac{\pi}{2}]\to\R$ is def\/ined by
\begin{gather*}
 \Upsilon_q^p
\left(
\begin{matrix}
n_{q}\\
l_{\alpha},l_{\beta}
\end{matrix};\vartheta_{q} \right)
:=\sqrt{\frac
{(2n_q\!+\alpha+\beta+1)(n_q\!+\alpha+\beta)! n_q!}
{(n_q\!+\alpha)! (n_q\!+\beta)!}}
(\cos\vartheta_q)^{l_\alpha}(\sin\vartheta_q)^{l_\beta}
P_{n_q}^{(\beta,\alpha)}(\cos(2\vartheta_q)),
\end{gather*}
with
$q=2^{j-1}+s-1$,
$l_\alpha=l_{2q}$, $l_\beta=l_{2q+1}$,
$\alpha=l_{2q}-1+2^{p-2-
\lfloor\log_{2} q\rfloor}$, and
$\beta=l_{2q+1}-1+2^{p-2-
\lfloor\log_{2} q\rfloor}$.
Note that we use the identif\/ication
\[
(l_{2^{p-1}},\ldots,l_{2^{p}-1})=(\left|m_{1}\right|,\ldots,\left|m_{2^{p-1}}\right|).
\]

\subsection*{Acknowledgements}

I~would like to thank A.F.M.~Tom ter Elst and Heather Macbeth for valuable
discussions.
I~would like to express my gratitude to the anonymous referees and an editor at
SIGMA whose helpful comments improved this paper.
Part of this work
was conducted while H.S.~Cohl was a~National Research Council Research
Postdoctoral Associate in the
Applied and Computational Mathematics Division
at the
National Institute of Standards and Technology, Gaithersburg, Maryland, USA.

\pdfbookmark[1]{References}{ref}
\LastPageEnding


\begin{thebibliography}{99}
\footnotesize\itemsep=0pt

\bibitem{Abra}
Abramowitz M., Stegun I.A., Handbook of mathematical functions with formulas,
  graphs, and mathematical tables, \textit{National Bureau of Standards Applied
  Mathematics Series}, Vol.~55, U.S. Government Printing Of\/f\/ice, Washington,
  D.C., 1964.

\bibitem{Izquierdoetal}
Alonso~Izquierdo A., Fuertes W.G., de~la Torre~Mayado M., Guilarte J.M.,
  One-loop corrections to the mass of self-dual semi-local planar topological
  solitons, \href{http://dx.doi.org/10.1016/j.nuclphysb.2007.11.023}{\textit{Nuclear Phys.~B}} \textbf{797} (2008), 431--463,
  \href{http://arxiv.org/abs/0707.4592}{arXiv:0707.4592}.

\bibitem{AAR}
Andrews G.E., Askey R., Roy R., Special functions, \textit{Encyclopedia of
  Mathematics and its Applications}, Vol.~71, Cambridge University Press,
  Cambridge, 1999.

\bibitem{Boyl}
Boyling J.B., Green's functions for polynomials in the {L}aplacian,
  \href{http://dx.doi.org/10.1007/BF00916651}{\textit{Z.~Angew. Math. Phys.}} \textbf{47} (1996), 485--492.


\bibitem{Cohlerratum12}
Cohl H.S., Erratum: Developments in determining the gravitational potential
  using toroidal functions, \href{http://dx.doi.org/10.1002/asna.201211723}{\textit{Astronom. Nachr.}} \textbf{333} (2012),
  784--785.

\bibitem{CohlthesisII}
Cohl H.S., Fourier and {G}egenbauer expansions for fundamental solutions of the
  {L}aplacian and powers in {${\mathbb R}^d$} and {${\mathbb H}^d$}, Ph.D.\
  thesis, The University of Auckland, Auckland, New Zealand, 2010.

\bibitem{Cohl12log}
Cohl H.S., Fourier expansions for a logarithmic fundamental solution of the
  polyharmonic equation, \href{http://arxiv.org/abs/1202.1811}{arXiv:1202.1811}.

\bibitem{CohlDominici}
Cohl H.S., Dominici D.E., Generalized {H}eine's identity for complex {F}ourier
  series of binomials, \href{http://dx.doi.org/10.1098/rspa.2010.0222}{\textit{Proc.~R. Soc. Lond. Ser.~A Math. Phys. Eng.
  Sci.}} \textbf{467} (2011), 333--345, \href{http://arxiv.org/abs/0912.0126}{arXiv:0912.0126}.

\bibitem{CohlKalII}
Cohl H.S., Kalnins E.G., Fourier and {G}egenbauer expansions for a fundamental
  solution of the {L}aplacian in the hyperboloid model of hyperbolic geometry,
  \href{http://dx.doi.org/10.1088/1751-8113/45/14/145206}{\textit{J.~Phys.~A: Math. Theor.}} \textbf{45} (2012), 145206, 32~pages,
  \href{http://arxiv.org/abs/1105.0386}{arXiv:1105.0386}.

\bibitem{CRTB}
Cohl H.S., Rau A.R.P., Tohline J.E., Browne D.A., Cazes J.E., Barnes E.I.,
  Useful alternative to the multipole expansion of $1/r$ potentials,
  \href{http://dx.doi.org/10.1103/PhysRevA.64.052509}{\textit{Phys. Rev.~A}} \textbf{64} (2001), 052509, 5~pages,
  \href{http://arxiv.org/abs/1104.1499}{arXiv:1104.1499}.

\bibitem{CTRS}
Cohl H.S., Tohline J.E., Rau A.R.P., Srivastava H.M., Developments in
  determining the gravitational potential using toroidal functions,
  \href{http://dx.doi.org/10.1002/1521-3994(200012)321:5/6<363::AID-ASNA363>3.0.CO;2-X}{\textit{Astronom. Nachr.}} \textbf{321} (2000), 363--372.

\bibitem{Cormenetal}
Cormen T.H., Leiserson C.E., Rivest R.L., Stein C., Introduction to algorithms,
  2nd ed., MIT Press, Cambridge, MA, 2001.

\bibitem{ErdelyiHTFII}
Erd\'{e}lyi A., Magnus W., Oberhettinger F., Tricomi F.G., Higher
  transcendental functions, Vol.~II, McGraw-Hill, New York~-- Toronto~--
  London, 1953.

\bibitem{FanoRau}
Fano U., Rau A.R.P., Symmetries in quantum physics, Academic Press Inc., San
  Diego, CA, 1996.

\bibitem{GelfandShilov}
Gel'fand I.M., Shilov G.E., Generalized functions. {V}ol.~{I}.~{P}roperties and
  operations, Academic Press, New York, 1964.

\bibitem{Grad}
Gradshteyn I.S., Ryzhik I.M., Table of integrals, series, and products, 7th
  ed., Elsevier/Academic Press, Amsterdam, 2007.

\bibitem{Ismail}
Ismail M.E.H., Classical and quantum orthogonal polynomials in one variable,
  \textit{Encyclopedia of Mathematics and its Applications}, Vol.~98, Cambridge
  University Press, Cambridge, 2005.

\bibitem{IPSWa}
Izmest'ev A.A., Pogosyan G.S., Sissakian A.N., Winternitz P., Contractions of
  {L}ie algebras and separation of variables. {T}he {$n$}-dimensional sphere,
  \href{http://dx.doi.org/10.1063/1.532820}{\textit{J.~Math. Phys.}} \textbf{40} (1999), 1549--1573.

\bibitem{IPSWb}
Izmest'ev A.A., Pogosyan G.S., Sissakian A.N., Winternitz P., Contractions of
  {L}ie algebras and the separation of variables: interbase expansions,
  \href{http://dx.doi.org/10.1088/0305-4470/34/3/314}{\textit{J.~Phys.~A: Math. Gen.}} \textbf{34} (2001), 521--554.

\bibitem{KilMS}
Kil'dyushov M.S., Hyperspherical functions of the ``tree'' type in the $n$-body
  problem, \textit{Soviet~J. Nuclear Phys.} \textbf{15} (1972), 113--118.

\bibitem{Koekoek1999}
Koekoek  J., Koekoek  R.,
 The Jacobi inversion formula,
 \href{http://dx.doi.org/10.1080/17476939908815177}{\textit{Complex Variables Theory Appl.}} {\bf 39}  (1999), 1--18,
 \href{http://arxiv.org/abs/math.CA/9908148}{math.CA/9908148}.

\bibitem{Lakeetal}
Lake M., Thomas S., Ward J., Non-topological cycloops, \href{http://dx.doi.org/10.1088/1475-7516/2010/01/026}{\textit{J.~Cosmol.
  Astropart. Phys.}} \textbf{2010} (2010), no.~1, 026, 27~pages,
  \href{http://arxiv.org/abs/0911.3118}{arXiv:0911.3118}.

\bibitem{Lin}
Lin C.D., Hyperspherical coordinate approach to atomic and other Coulombic
  three-body systems, \href{http://dx.doi.org/10.1016/0370-1573(94)00094-J}{\textit{Phys. Rep.}} \textbf{257} (1995), 1--83.

\bibitem{LinYang}
Lin F., Yang Y., Energy splitting, substantial inequality, and minimization for
  the {F}addeev and {S}kyrme models, \href{http://dx.doi.org/10.1007/s00220-006-0123-0}{\textit{Comm. Math. Phys.}} \textbf{269}
  (2007), 137--152.

\bibitem{MOS}
Magnus W., Oberhettinger F., Soni R.P., Formulas and theorems for the special
  functions of mathematical physics, 3rd ed., \textit{Die Grundlehren der
  mathematischen Wissenschaften}, Bd.~52, Springer-Verlag, New York, 1966.

\bibitem{Miller}
Miller Jr. W., Symmetry and separation of variables, \textit{Encyclopedia of Mathematics and its Applications},
Vol.~4, Addison-Wesley Publishing Co., Reading,
Mass.~-- London~-- Amsterdam, 1977.

\bibitem{NIST}
Olver F.W.J., Lozier D.W., Boisvert R.F., Clark C.W. (Editors), N{IST} handbook
  of mathematical functions, U.S. Department of Commerce National Institute of
  Standards and Technology, Washington, DC, 2010.

\bibitem{Schw}
Schwartz L., Th\'eorie des distributions. {T}ome~{I}, Actualit\'es Sci. Ind.,
  no. 1091, Hermann \& Cie., Paris, 1950.

\bibitem{Sloane}
Sloane N.J.A., Plouf\/fe S., The encyclopedia of integer sequences, Academic
  Press Inc., San Diego, CA, 1995.

\bibitem{Stanley}
Stanley R.P., Enumerative combinatorics. {V}ol.~2, \textit{Cambridge Studies in
  Advanced Mathematics}, Vol.~62, Cambridge University Press, Cambridge, 1999.

\bibitem{Szego}
Szeg{\H{o}} G., Orthogonal polynomials, \textit{American Mathematical Society
  Colloquium Publications}, Vol.~23, Amer. Math. Soc., Providence, R.I., 1959.

\bibitem{Vilen}
Vilenkin N.Ja., Special functions and the theory of group representations,
  \textit{Translations of Mathematical Monographs}, Vol.~22, Amer. Math. Soc.,
  Providence, R.I., 1968.

\bibitem{VilenkinKlimyk2}
Vilenkin N.Ja., Klimyk A.U., Representation of {L}ie groups and special
  functions. {V}ol.~2.~Class I representations, special functions, and integral
  transforms, \textit{Mathematics and its Applications (Soviet Series)},
  Vol.~74, Kluwer Academic Publishers Group, Dordrecht, 1993.

\bibitem{VilKuzSmor}
Vilenkin N.Ja., Kuznetsov G.I., Smorodinski{\u\i} Ya.A., Eigenfunctions of the
  {L}aplace operator providing representations of the {${\rm U}(2)$}, {${\rm
  SU}(2)$}, {${\rm SO}(3)$}, {${\rm U}(3)$} and {${\rm SU}(3)$} groups and the
  symbolic method, \textit{Soviet~J. Nuclear Phys.} \textbf{2} (1965),
  645--652.

\bibitem{WenAvery}
Wen Z.Y., Avery J., Some properties of hyperspherical harmonics,
  \href{http://dx.doi.org/10.1063/1.526621}{\textit{J.~Math. Phys.}} \textbf{26} (1985), 396--403.

\end{thebibliography}
\end{document}